\documentclass[12pt]{article}
\usepackage{epsfig}
\newcommand{\be}{\begin{equation}}
\newcommand{\ee}{\end{equation}}
\newcommand{\lee}[1]{\label{#1} \end{equation}}
\newcommand{\bea}{\begin{eqnarray}}
\newcommand{\leea}[1]{\label{#1} \end{eqnarray}}
\newcommand{\eea}{\end{eqnarray}}
\newcommand{\nn}{\nonumber}
\newcommand{\eq}[1]{eq.~(\ref{#1})}
\newcommand{\Eq}[1]{Eq.~(\ref{#1})}
\newcommand{\eqs}[2]{eqs.~(\ref{#1}) and (\ref{#2})}

\newcommand{\fig}[1]{fig.~(\ref{#1})}
\newcommand{\loadeps}[1]{\epsfig{file=#1.eps, scale=1.0, width=45mm}}
\newcommand{\diagform}[2]{\put(60,#1){\makebox(110,44.61)[l]{
  \begin{minipage}{110mm} #2 \end{minipage} }} }
\newcommand{\al}{\alpha}

\newcommand{\ga}{\gamma}
\newcommand{\de}{\delta}

\newcommand{\varep}{\varepsilon}

\newcommand{\et}{\eta}

\newcommand{\la}{\lambda}

\newcommand{\si}{\sigma}

\renewcommand{\th}{\theta}

\newcommand{\ch}{\chi}

\newcommand{\om}{\omega}

\newcommand{\Ga}{\Gamma}
\newcommand{\De}{\Delta}
\newcommand{\La}{\Lambda}
\newcommand{\Si}{\Sigma}

\newcommand{\Om}{\Omega}
\def\bbbone{{\mathchoice {\rm 1\mskip-4mu l} {\rm 1\mskip-4mu l}
 {\rm 1\mskip-4.5mu l} {\rm 1\mskip-5mu l}}}

\begin{document}
\title{Exact renormalization group analysis \\
in Hamiltonian theory: \\
I. QED Hamiltonian on the light front \vspace*{.5cm}}
\author{E.~L.~Gubankova\thanks{e-mail: elena@hal5000.tphys.uni-heidelberg.de},
 F.~Wegner \vspace*{.5cm}\\
\normalsize\it Institut f\"ur Theoretische Physik der Universit\"at Heidelberg \\
\normalsize\it Philosophenweg 19, D69120 Heidelberg, FRG}
\date{Februay 22, 1997}

\maketitle

\vspace*{1cm}
\begin{abstract}
The infinitesimal unitary transformation, introduced recently by
F.~Wegner, to bring the Hamiltonian to diagonal (or band diagonal)
form, is applied to the Hamiltonian theory as an exact renormalization 
scheme. We consider QED on the light front to illustrate
the method. The low-energy generated interaction, induced in the renormalized 
Hamiltonian to the order $\alpha$, is shown to be negative to
insure, together with instantaneous term and perturbative photon exchange,
the bound states for positronium. It is possible to perform the
complete elimination of the $ee\ga$-vertex in the instant form frame;
this gives rise to the cutoff independent $e\bar{e}$-interaction
governed by generated and instantaneous terms. The well known
result for the singlet-triplet splitting $\frac{7}{6} \alpha^2 \rm{Ryd}$ 
is recovered in the nonrelativistic limit as long as $\la <\!\!< m$.

We examine the mass and wave function renormalization.
The ultraviolet divergencies, associated with a large transverse momentum,
are regularized by the regulator arising from the unitary transformation.
The severe infrared divergencies are removed if all diagrams to the second
order, arising from flow equations method and normal-ordering Hamiltonian, 
are taken into account. The electron (photon) mass in the renormalized
Hamiltonian vary with UV cutoff in accordance with $1$-loop renormalization
group equations. This indicates an intimate connection between Wilson's
renormalization and the flow equation method.

The advantages of the method in comparison with the naive renormalisation
group approach are discussed.
\end{abstract}

\newpage
\section{Introduction}

We apply the method formulated by F.~Wegner \cite{We} of unitary transformation,
to bring the Hamiltonian to diagonal (or near diagonal) form, to the problem
of renormalization. We start with the regularized bare cutoff Hamiltonian, 
with the energy widths (the energy differences between the free states 
in a matrix element) restricted to be below the bare UV cutoff $\Lambda$. 
As renormalization step we perform the unitary transformation 
that removes the energy widths in 
the final Hamiltonian, to be below the final cutoff $\lambda < \Lambda$,
resulting in the band-diagonal structure for the renormalized Hamiltonian.
 This corresponds to the Wilson renormalization
procedure in the Lagrange theory of integrating out the fields of
higher energies. 

This unitary transformation can be written
\be
U(\la,\La) \; H_B(\La, e_{\La}, m_{\La}, g(e_{\La}, m_{\La})) \; U^+(\la,\La) =
H_B(\la, e_{\la}, m_{\la}, g(e_{\la}, m_{\la}))
\; , \lee{in1}
where $e_{\la}, m_{\la}$ denote the set of finite number of independent
coupling constants, in the case of QED - the $ee\ga$-coupling constant
and the fermion mass.
Instead we generate in the renormalized Hamiltonian new 
effective interactions, denoted as ${g(e_{\la,}m_{\la})}$, and 
corresponding to the operators of higher dimensions. They must be added
to the bare cutoff theory as irrelevant operators, namely
\be
g(e_{\La}, m_{\La}) = 0 \, ,\quad \La\rightarrow \infty
\; , \lee{in2}
being manifest at the low-energy scale. The number of independent coupling
constants of the theory is preserved, i.e. $g\rightarrow 0$
when $e\rightarrow 0, m\rightarrow 0$.

By definition the renormalized Hamiltonian, see \cite{GlWi}, in the literature also
called the effective Hamiltonian, is
\be
H_R = \lim_{\La\to\infty} H_B
\; . \lee{in3}

For renormalizable theories there arise only a finite number of infinities,
associated with the limit $\La \to \infty$.
They can be eliminated order by order in the running coupling 
$e_{\lambda}$ by an appropriate choice of cutoff-dependent 
counterterms. The counterterms are chosen from the constraint
on the renormalized Hamiltonian to run coherently with the cutoff.

This renormalization scheme, called similarity Hamiltonian approach,
was presented by S.~Gla\-zek and K.~Wil\-son \cite{GlWi} as an alternative to the
traditional Lagrange method.

The similarity unitary transformation was used by these authers 
in light-front field theory to obtain the band-diagonal Hamiltonian
with finite cutoff $\lambda$. For these purposes we exploit 
in this work the flow equations, formulated by Wegner \cite{We}, that perform
infinitesimal unitary transformations in a continuous way. 

We consider in this work two aspects of renormalized Hamiltonian. The first
point is related with renormalization group analysis. The second one is
connected with the low energy sector of the theory and bound state calculations.

In the next section we give the key ingredients of the flow equation method 
and discuss and its application to renormalization group analysis in Hamiltonian theory.
 
\section{Flow equations}

The flow equation is written 
\be
\frac{dH(l)}{dl} = [\eta(l),H(l)]
\lee{fe1}
or for the matrix elements
\be
\frac{dH_{ij}}{dl}=\eta _{ij}(E_j-E_i)+[\eta,H_I]_{ij}
\; , \lee{fe2}
with the Hamiltonian $H = H_0 + H_I$ devided into a free and an interacting part.
Here $\eta (l)$ is the generator of the unitary transformation,
that determines in the leading order the evolution of the interacting part
$H_{I,ik},i \neq k$ or, vice verca, can be regarded as function of $H_I$,
i.e. $\eta (H_I)$. The flow
parameter $l$ changes from $l_{\La} = 0$ as $\La\rightarrow\infty$,
that corresponds to the initial
canonical Hamiltonian $H_{can}$,
to some finite value $l_{\la}$, where the Hamiltonian has 
band diagonal form.
The value $l_{\la} = \infty$ corresponds to the diagonal (or block-diagonal,
in the case when exact diagonalization is impossible) form of the Hamiltonian.
The dimension of the flow parameter $l$ is $1/({\rm energy})^2$.

In the problem of renormalization the physical sense of the UV cutoff 
can be assigned to the flow parameter $l$, explicitly it satisfies
\be
l_{\la} = \frac{1}{\lambda ^2}
\; , \lee{fe3}
where $\la$ is the UV cutoff.
The unitary transformation removes the matrix elements between the free
states with energy differences between the bare cutoff, $\Lambda$,
and the final cutoff, $\lambda$ ($\lambda < \Lambda$), resulting in a
band-diagonal form for $H_{\lambda}$.

We continue with the method of flow equations itself. The generator
of the transformation, according to \cite{We}, is chosen
\be
\eta_{ij} = [H_0,H]_{ij} = (E_i-E_j) H_{I,ij}
\lee{fe4}
giving
\be
\frac{dH_{ij}}{dl} = -(E_i-E_j)^2~H_{I,ij} + [\eta,H_I]_{ij}
\; . \lee{fe5}

Generaly the interacting part is associated with the coupling constant $g$.
We proceed further in the frame of perturbative theory (PT) with respect
to the interaction $H_I$, or therefore O(g). From the equations above follows, that
the generator $\eta$ to the leading order $O(H_I)$
generates the effective interaction 
to the order $O(H_I^2)$. The new term must be added to the initial
Hamiltonian as irrelevant operator at $l_{\La} \rightarrow 0$ 
(i.e. vanishing at $l=0$, $H_{gen}(l=0)=0$) but being nonzero 
for any finite value of $l_{\la}$. To make the
effective interaction band-diagonal we need to introduce the generator
to the next order $O(H_I^2)$, which in its turn generates the 
effective interaction to the third order $O(H_I^3)$.
We truncate this series assuming the coupling constant to be small.
Therefore for finite value of $l_{\la}$ one has 
\bea
H(l) &=& H_0(l)+H^{(1)}(l)+H^{(2)}(l) + \ldots \nn \\
\eta(l) &=& \eta^{(1)}(l)+\eta^{(2)}(l) + \ldots
\; , \leea{fe6}
where the upper index denotes the order of PT, and 
\mbox{$H^{(1)}(l) = H_I(l)$}, with \mbox{$\lim_{\La \rightarrow \infty} H_{I}(l_{\La}=0)$}, 
being the interaction present in the initial canonical Hamiltonian.

To the leading order of perturbative theory (PT)
\be
\frac{dH_{ij}^{(1)}}{dl} = -(E_i-E_j)^2 H_{ij}^{(1)}
\; . \lee{fe7}
Neglecting the dependence of the energies $E_i$ on the flow parameter
(being of higher order), we obtain the solution in the form
\bea
&& H_{ij}^{(1)}(l_{\la})=H_{I,ij}(l_{\la})=
H_{I,ij}(l_{\La}) \frac{f_{ij}(l_{\la})}{f_{ij}(l_{\La})}
\; , \leea{fe8}
where we have intoduced
\bea
f_{ij}(l_{\la}) = \exp\left\{ -l_{\la}(E_i-E_j)^2 \right\}
= \exp\left\{-\left(\frac{E_i-E_j}{\la}\right)^2 \right\}
\; . \leea{fe9}
The initial condition is defined at the bare cutoff $\La\rightarrow\infty$, 
namely $\lim_{\La\rightarrow\infty}H_{I}(l_{\La})$ 
corresponds in the case of QED to the bare coupling constant $e_0$.

The structure of the expression for $H_I(l_{\la})$ is transparent.
By the unitary transformation performed with the leading order generator 
$\eta$, we succeded to eliminate far-off-diagonal elements from the 
Hamiltonian matrix, so that only matrix elements with
energy differences $|E_i-E_j|<\frac{1}{\sqrt{l}}=\la$ are present.

To the next to leading order $O({H_I}^2)$
\be
\frac{dH_{ij}^{(2)}}{dl} = -(E_i-E_j)^2H_{ij}^{(2)}
+[\eta ^{(1)},H^{(1)}]_{ij}
\; . \lee{fe10}
We introduce $H_{ij}^{(2)}(l_{\la}) = f_{ij}(l_{\la}) 
\tilde{H}_{ij}^{(2)}(l_{\la})$, then the flow equation is rewritten
(again neglecting the cutoff dependence of the energies $E_i$)
\be
\frac{d\tilde{H}_{ij}^{(2)}}{dl}=\frac{1}{f_{ij}} [\eta ^{(1)},\\
H^{(1)}]_{ij}
\; . \lee{fe11}
This gives rise to the following solution
\bea
&& \tilde{H}_{ij}^{(2)}(l_{\la})=\tilde{H}_{ij}^{(2)}(l_{\La})
+\int_{l_{\La}}^{l_{\la}}dl'\frac{1}{f_{ij}} [\eta ^{(1)},H^{(1)}]_{ij}
\; . \leea{fe12}
We are not going to give further details. But it is worth to mention here, 
that in the case of QED the second order term $H^{(2)}$ contributes 
(in different sectors) to the new generated interaction $H^{(gen)}$
and to the electron (photon) self energy terms.
This means that beginning from the second order the values
$\tilde{H}_{ij}^{(n)}(l_{\la})$, defined as
$H_{ij}^{(n)}(l_{\la}) = f_{ij}(l_{\la}) \, \tilde{H}_{ij}^{(n)}(l_{\la})$,
contain two terms, the interaction part and counterterms. Generally it can
be written
\be
\tilde{H}^{(n)}(l_{\la}) = \tilde{H}_I^{(n)}(l_{\la})
 + X^{(n)}(l_{\la})  
\; . \ee
The counterterms are determined from the coupling coherence condition,
\cite{GlWi}. For the explicit treatment of the second order for QED, see later. 

Generally, the system of self-consistent flow equations
for the Hamiltonian $H(l)$ and the generator of the unitary 
transformation $\eta (l)$ can be written in matrix form
as follows 
\bea
&& \frac{dH_{ij}(l)}{dl} = [\eta ,H_I]_{ij}+
\frac{dlnf_{ij}}{dl}H_{ij}\nn\\
&& \eta_{ij}(l)=\frac{1}{E_i-E_j}\left(-\frac{dlnf_{ij}}{dl}H_{ij} \right)
\; , \leea{fe13}
where the function $f_{ij}$ determines how fast the non-diagonal
part of the Hamiltonian matrix with $|E_i-E_j|> \lambda$ vanishes with $l$,
$f_{ij} = \exp(-l(E_i-E_j)^2)$
Note, that this form of the function $f_{ij}$ corresponds to 
the choice $\eta=[H_0,H]$ for the generator.
Other choices for the function $f_{ij}(l)$ are possible (see Appendix A).

The structure of the flow equations, \eq{fe13},  is transparent. The first
term in \eq{fe13} is responsible for the renormalization of physical
values (coupling constants, masses, wave functions) and also for
the structure of the new generated interactions to the higher order
(with respect to the canonical interaction) being then present
in the initial Hamiltonian; the second term in \eq{fe13} insures
band diagonal structure of the renormalized Hamiltonian.
As will be shown later, in the case of QED the physical masses
run to the order $g^2$, while the coupling constant starts to run
to the order $g^3$ with the right renormalization group coefficients,
i.e. with those obtained in standard perturbative theory.

The second equation for $\eta$ is chosen in the form that the terms with
small energy denominators $(|E_i-E_j| \rightarrow 0)$ effectively
do not contribute in the renormalized Hamiltonian. This advantage,
as compared with naive perturbative theory, enables to consider
Hamiltonians with continuum spectrum. We mention that the problem of small
energy denominators is solved also in other schemes (Appendix A).

\section{Renormalized Hamiltonian \boldmath$H_R$ to the second order}

\subsection{Canonical light-front \boldmath$QED_{3+1}$ Hamiltonian}

We start with the canonical light-front QED Hamiltonian $H_{can}$,
devided into free and interacting parts 
\be
P^-=H_{can} = \int dx^-d^2x^{\bot}({\cal H}_0+{\cal H}_I)
\; . \lee{ch1}
In light-front gauge $A^+=A^0+A^3=0$, the constrained degrees of freedom
$A^-$ and $\psi _-$ ($\psi=\psi _+ +\psi_- ,\psi_{\pm}=\Lambda_{\pm}\psi$)
can be removed explicitly; this gives the light-front gauge Hamiltonian
defined through the independent physical fields 
$A_{\bot}$ and $\psi_+$ only \cite{ZhHa}
\be
{\cal H}_0=\frac{1}{2}(\partial ^iA^j)(\partial ^iA^j)+\\
\xi^+ \left( \frac{-\partial _{\bot}^2+m^2}{i\partial ^+} \right) \xi
\; , \lee{ch2}
\be
{\cal H}_I
 = {\cal H}_{ee \gamma} + {\cal H}_{ee \gamma \gamma} + {\cal H}_{eeee}
\lee{ch3}
and 
\be
{\cal H}_{ee \gamma}=e\xi^+ \left[ -2(\frac{\partial ^{\bot}}{\partial ^+}
\cdot A^{\bot})+\sigma \cdot A^{\bot}\frac{\sigma \cdot\partial ^{\bot}+
m}{\partial ^+}+\frac{\sigma \cdot \partial ^{\bot}+m}{\partial ^+}
\sigma \cdot A^{\bot} \right] \xi
\; , \lee{ch4}
\be
{\cal H}_{ee \gamma \gamma}= -ie^2 \left[ \xi ^+ \sigma\cdot A^{\bot}
\frac{1}{\partial ^+}(\sigma \cdot A^{\bot}\xi) \right]
\; , \lee{ch5}
\be
{\cal H}_{eeee}=2e^2 \left[ \left( \frac{1}{\partial ^+}(\xi ^+ \xi) \right)
 \left( \frac{1}{\partial ^+}(\xi ^+ \xi) \right) \right]
\; , \lee{ch6}
where $\{\sigma ^i\}$ are the standard $2 \times 2$ Pauli matrices, and
$\partial ^+=2\partial _- = 2\frac{\partial}{\partial x^-}$.
We have used the two-component representation for fermion fields introduced
by Zhang and Harindranath \cite{ZhHa} $\psi_+={\xi \choose 0}$.
To simplify the calculations we rewrite all interactions through creation
and annihilation operators. This turns out to be useful in the flow equations
formalism, \cite{Mi}.

Following standard quantum field theory procedure
we use the momentum-space representation for the field operators, \cite{PeWi}
and \cite{ZhHa},
\bea
\xi(x) &=& \sum_s \chi_s \int\frac{dp^+d^2p^{\bot}}{2(2\pi)^3}\theta(p^+)
(b_{p,s}e^{-ipx}+d_{p,\bar{s}}e^{ipx}) \nn \\
A^i(x) &=& \sum_{\lambda}\int \frac{dq^+ d^2q^{\bot}}{2(2\pi)^3}
\frac{\theta (q^+)}{\sqrt{q^+}}(\varepsilon _{\lambda}^i a_{q,\lambda}
e^{-iqx}+h.c.)
\; , \leea{ch8}
where spinors are $\chi _{1/2}^{tr}=(1,0)$, $\chi_{{-1/2}^{tr}}=(0,1)$,
with $\bar{s}=-s$
and polarization vectors $\varepsilon_1^i=\frac{-1}{\sqrt{2}}(1,i)$,
$\varepsilon_{-1}^i=\frac{1}{\sqrt{2}}(1,-i)$;
the integration running over the $p^+\ge 0$ only these
states, that are allowed the light-front theory.  

The corresponding (anti)commutation relations are
\bea
& \{ b_{p,s},b_{p',s'}^+\}=\{d_{p,s},d_{p',s'}^+\}=\bar{\de}_{p,p'}\delta_{ss'} & \nn \\
& [a_{q,\lambda},a_{q',\lambda '}^+]=\bar{\de}_{q,q'} \delta_{\lambda,\lambda '} &
\; , \leea{ch10}
where 
\be
\bar{\de}_{p,p'}\equiv 2(2\pi)^3\delta(p^+-p'^+)\delta^{(2)}
(p^{\bot}-p'^{\bot}) 
\; . \lee{ch11}
The light-front vacuum has trivial structure for both boson and fermion
sectors, namely $a_q|0>=0$; $b_p|0>=0$, simpifying the 
analitical calculations.
The normalization of states is according to
\be
<p_1,s_1|p_2,s_2>=\bar{\de}_{p_1,p_2}\delta_{s_1,s_2}
\; , \lee{ch12}
where $b_{p,s}^+|0>=|p,s>$.

Making use of the field representation \eq{ch8},
we have the following Fourier transformed for

\noindent
the {\bf free} Hamiltonian 
\be
H_0=\sum _s\int\frac{dp^+ d^2p^{\bot}}{2(2\pi)^3}\theta (p^+)\\
\frac{p^{\bot 2}+m^2}{p^+} (b_{p,s}^+b_{p,s}+d_{p,s}^+d_{p,s})+\\
\sum_{\lambda}\int\frac{dq^+d^2q^{\bot}}{2(2\pi)^3}\theta(q^+)\\
\frac{q^{\bot 2}}{q^+}a_{q,\lambda}^+a_{q,\lambda}
\; , \lee{ch13}

\noindent
the leading order $O(e)$ {\bf \boldmath$ee\gamma$-coupling} 
\bea
H_{ee\gamma}&=&\sum_{\lambda s_1s_2}\int_{p_1p_2q} \!\!\!
  [g_{p_1p_2q}^*(l) \varepsilon_{\lambda}^i\tilde{a}_{q}
 + g_{p_1p_2q}(l) \varepsilon_{\lambda}^{i *}\tilde{a}_{-q}^+]
(\tilde{b}_{p_2}^+\tilde{b}_{p_1} +\tilde{b}_{p_2}^+\tilde{d}_{-p_1}^+ +
\tilde{d}_{-p_2}\tilde{b}_{p_1} +\tilde{d}_{-p_2}\tilde{d}_{-p_1}^+)\nonumber\\
& &\times\chi_{s_2}^+\Gamma_l^i(p_1,p_2,-q)\chi_{s_1} \bar{\de}_{q,p_2-p_1}
\; , \leea{ch14}
where
\be
\Gamma_l^i(p_1,p_2,q)=2\frac{q^i}{q^+}-\\
\frac{\sigma\cdot p_2^{\bot}-im}{p_2^+}\sigma^i-\\
\sigma^i\frac{\sigma\cdot p_1^{\bot}+im}{p_1^+}
\; , \lee{ch15}
where the mass is l-dependent.
Further we have for the {\bf instantaneous} interactions of the order $O(e^2)$
\bea
H_{eeee}^{inst}&=&\sum_{s_1s_2s_3s_4}\int_{p_1p_2p_3p_4}
\!\!\!\!\!g_{p_1p_2p_3p_4}^{eeee}(l)
(\tilde{b}_{p_3}^+ +\tilde{d}_{-p_3})(\tilde{b}_{p_4}^+ +\tilde{d}_{-p_4})
(\tilde{b}_{p_1} +\tilde{d}_{-p_1}^+)(\tilde{b}_{p_2} +\tilde{d}_{-p_2}^+)
\nonumber \\
& &\times\chi_{s_3}^+\chi_{s_4}^+\frac{4}{(p_1^+-p_3^+)^2}\chi_{s_1}\chi_{s_2}
\bar{\de}_{p_3+p_4,p_1+p_2}
\leea{ch16}
and
\bea
H_{ee\gamma\gamma}^{inst}&=&\sum_{s_1s_2\lambda_1\lambda_2}
\int_{p_1p_2q_1q_2}\!\!\!\!\!g_{p_1p_2q_1q_2}^{ee\gamma \gamma}(l)
(\varepsilon_{\lambda_1}^{i *}\tilde{a}_{q_1}^+ +
\varepsilon_{\lambda_1}^i\tilde{a}_{-q_1})
(\varepsilon_{\lambda_2}^j\tilde{a}_{q_2}+
\varepsilon_{\lambda_2}^{j *}\tilde{a}_{-q_2}^+)
(\tilde{b}_{p_2}^+ +\tilde{d}_{-p_2})(\tilde{b}_{p_1}+\tilde{d}_{-p_1}^+)
\nonumber\\
& &\times\chi_{s_2}^+\frac{\sigma^j\sigma^i}{(p_1^+-q_1^+)}\chi_{s_1}
\bar{\de}_{p_1+q_2,q_1+p_2}
\; ; \leea{ch17}
here 
\bea
& \tilde{a}_q\equiv a_{q,\lambda}\frac{\theta(q^+)}{\sqrt{q^+}}, \qquad
 \left[ \tilde{a}_{-q}\equiv a_{-q,\lambda}\frac{\theta(-q^+)}{\sqrt{-q^+}} \right]
 \; , & \nn \\
& \tilde{b}_p\equiv b_{p,s}\theta(p^+), \qquad
 \tilde{d}_p\equiv d_{p,\bar{s}}\theta(p^+)
\; , \leea{ch19}
and the $\bar{\de}$ stands for the function defined in \eq{ch11}, the
short notation for the integral we understand as
\be
\int_p\equiv\int\frac{dp^+d^2p^{\bot}}{2(2\pi)^3}
\; . \lee{ch20}

In the formulas above we write explicitly the momentum dependence of 
the coupling constants as long as $l\neq 0$. The initial conditions 
for the couplings are defined at the value of the bare cutoff 
$\La\rightarrow\infty$, namely   
\be
\lim_{\La\rightarrow\infty}g^{ee\gamma}(l_{\La})=e_0
\lee{ch21}
and for both instantaneous interaction couplings
\be 
\lim_{\La\rightarrow\infty}g^{inst}(l_{\La})=e_0^2
\; ; \lee{ch22}
these correspond to the couplings of the canonical theory.

\subsection{The flow equations in \boldmath$|e\bar{e}\!>$-sector} 

\subsubsection{Generated interaction}

Following the procedure outlined in the second section, the leading order
generator of the unitary transformation is
\bea
\eta^{(1)}(l)&=&\sum_{\lambda s_1s_2}\int_{p_1p_2q}\!\!\!(\eta_{p_ip_f}^*(l)
\varepsilon_{\lambda}^i\tilde{a}_q+
\eta_{p_ip_f}(l)\varepsilon_{\lambda}^{i *}\tilde{a}_{-q}^+)
(\tilde{b}_{p_2}^+\tilde{b}_{p_1}+\tilde{b}_{p_2}^+\tilde{d}_{-p_1}^+ +
\tilde{d}_{-p_2}\tilde{b}_{p_1}+\tilde{d}_{-p_2}\tilde{d}_{-p_1}^+)\nonumber\\
& &\times\chi_{s_2}^+\Gamma_l^i(p_1,p_2,-q)\chi_{s_1} \bar{\de}_{q,p_2-p_1}
\; , \leea{gi1}
where $p_i$ and $p_f$ stand for the set of initial and final momenta,
respectively, and
\be
\eta_{p_ip_f}(l)=-\Delta_{p_ip_f}g_{p_ip_f}=\\
\frac{1}{\Delta_{p_ip_f}}\cdot\frac{dg_{p_ip_f}}{dl}
\; . \lee{gi2}

Further we calculate the bound states of positronium.
In what follows consider in $|e\bar{e}>$ sector

\noindent
the {\bf generated interaction} to the first nonvanishing order
\be
H_{e\bar{e}e\bar{e}}^{gen}=\sum_{s_1\bar{s}_2s_3\bar{s}_4}\\
\int_{p_1p_2p_3p_4}\\
V_{p_ip_f}^{gen}(l)b_{p_3}^+d_{p_4}^+d_{p_2}b_{p_1}\\
\chi_{s_3}^+\chi_{\bar{s}_4}^+\chi_{\bar{s}_2}\chi_{s_1}\\
\bar{\de}_{p_1+p_2,p_3+p_4}
\; , \lee{gi3}
with the initial condition 
$\lim_{\La\rightarrow\infty}V_{p_ip_f}^{gen}(l_{\La})=0$,

\noindent
and the {\bf instantaneous interaction}
\be
H_{e\bar{e}e\bar{e}}^{inst}=\sum_{s_1\bar{s}_2s_3\bar{s}_4}\\
\int_{p_1p_2p_3p_4}\\
V_{p_ip_f}^{inst}(l)b_{p_3}^+d_{p_4}^+d_{p_2}b_{p_1}\\
\chi_{s_3}^+\chi_{\bar{s}_4}^+\chi_{\bar{s}_2}\chi_{s_1}\\
\bar{\de}_{p_1+p_2,p_3+p_4}
\; , \lee{gi4}
where 
\be
V_{p_ip_f}^{inst}(l) = g_{p_ip_f}^{inst}(l) \, \frac{4}{(p_1^+-p_3^+)^2}(l)
\; . \lee{gi5}
The ordering of the field operators in both interactions
is important because it defines the kind of interaction,
attractive or repulsive; the ordering given
satisfies the standard Feynmann rule prescription in the
$|e\bar{e}>$ sector.

We neglect the $l$ dependence of momenta in the interaction, which
enables us to write the flow equations for the corresponding couplings.
 
The flow equations to the order $O(e^2)$ 
\bea
\frac{dg_{p_ip_f}(l)}{dl}&=&-\Delta_{p_ip_f}^2g_{p_ip_f}(l)\nonumber\\
\frac{dg_{p_ip_f}^{inst}(l)}{dl}&=&-\Delta_{p_ip_f}^2g_{p_ip_f}^{inst}(l)\\
\frac{dV_{p_ip_f}^{gen}(l)}{dl}&=&<[\eta^{(1)}(l),H_{ee\gamma}]>_{|e\bar{e}>}
-\Delta_{p_ip_f}^2V_{p_ip_f}^{gen}(l)\nonumber
\; , \leea{gi6}
where
\be
\Delta_{p_ip_f} = \sum p_i^- - \sum p_f^-
\lee{gi7}
and the light-front fermion energie is \mbox{$p^- = \frac{p^{\bot 2} + m^2}{p^+}$},
the photon one \mbox{$q^- = \frac{q^{\bot 2}}{q^+}$}.
The matrix element \mbox{$<[\eta^{(1)}(l) , H_{ee\gamma}]>_{|e\bar{e}>}$}
is understood as the corresponding commutator between
the free electron-positron states, namely
\mbox{$<p_3 s_3, p_4 \bar{s}_4|...|p_1 s_1, p_2 \bar{s}_2>$}.   

Neglecting the dependence of the momenta (the fermion mass) on 
the flow parameter $l$, the solution reads 
\bea
g_{p_ip_f}(l)&=&f_{p_ip_f}\cdot e_0+O(e^3)\nonumber\\
g_{p_ip_f}^{inst}(l)&=&f_{p_ip_f}\cdot e_0^2+O(e^4)\\
V_{p_ip_f}^{gen}(l)&=&f_{p_ip_f}\cdot\int_0^l dl'
\frac{1}{f_{p_ip_f}(l')}<[\eta^{(1)}(l'),H_{ee\gamma}(l')]>_{|e\bar{e}>}
\nonumber
\; , \leea{gi8}
where the subscript indicates, that the commutator is considered 
in the electron-positron sector. As was discussed in the previous section,
the function $f_{p_ip_f}$ \eq{fe9} insures band-diagonal structure for the
renormalized interaction.

The matrix element  of the commutator $[\eta^{(1)},H_{ee\gamma}]$
in the exchange and annihilation channels is (Appendix B)
\be
<[\eta^{(1)},H_{ee\gamma}]>/\delta_{p_1+p_2,p_3+p_4} = 
\left\{ \begin{array}{l}
 M_{2ii}^{ex}  \frac{1}{(p_1^+-p_3^+)}(\eta_{p_1,p_3}g_{p_4,p_2}+
\eta_{p_4,p_2}g_{p_1,p_3}) \; , \\
\\
-M_{2ii}^{an}  \frac{1}{(p_1^++p_2^+)}(\eta_{p_1,-p_2}g_{p_4,-p_3}+
\eta_{p_4,-p_3}g_{p_1,-p_2}) \; ,
\end{array} \right. 
\lee{gi9}
\noindent
where
\bea
&& \eta_{p_1,p_2} = e_0\cdot\frac{1}{\Delta_{p_1p_2}}
 \frac{df_{p_1,p_2}}{dl} \nn \\
&&g_{p_1,p_2} = e_0\cdot f_{p_1,p_2}
\leea{gi10}
and $\Delta_{p_1,p_2} = p_1^- - p_2^- - (p_1-p_2)^-$.
In what follows we drop the index $\,0\,$ at the coupling.
The matrix elements $M_{2ii}$ between the corresponding spinors
in both channels
\bea
M_{2ij}^{(ex)}&=&[\chi_{s_3}^+\Gamma_l^i(p_1,p_3,p_1-p_3)\chi_{s_1}]\,
[\chi_{\bar{s}_2}^+\Gamma_l^j(-p_4,-p_2,-(p_1-p_3))\chi_{\bar{s}_4}]
\nonumber\\
\\
M_{2ij}^{(an)}&=&[\chi_{s_3}^+\Gamma_l^i(-p_4,p_3,-(p_1+p_2))\chi_{\bar{s}_4}] \,
[\chi_{\bar{s}_2}^+\Gamma_l^j(p_1,-p_2,p_1+p_2)\chi_{s_1}]\nonumber
\leea{gi11}
determine the spin structure of the effective interaction.
It is calculated explicitly in light-front frame in Appendix B.

The form of the second order renormalized interactions \eq{gi8},
expressed through the $f$-function (that defines the behaviour
of the first order coupling), is universal for the different 
renormalization schemes (see Appendix). (This is true up
to some factor in the integral for the generated interaction). Specifying
the $f$ function we obtain the explicit form of the renormalized
interactions for the different unitary transformations.
In Appendix C we compare the results for the second order generated
interaction in two renormalization schemes.

Here we choose the $f$-function as
\be
f_{p_i,p_f} = \exp(-l \, \Delta_{p_i,p_f}^2)
\; . \lee{gi12}
    
Then, neglecting again the dependence of the momenta on the flow parameter $l$,
we have for the generated interaction in both channels 
\bea
V^{(ex)}(l)&=&-e^2M_{2ii}^{(ex)}\frac{1}{(p_1^+-p_3^+)}
(\Delta_{p_1,p_3}+\Delta_{p_4,p_2})f_{p_ip_f}\cdot
\int_0^l dl'\frac{f_{p_1,p_3}(l')f_{p_4,p_2}(l')}{f_{p_ip_f}(l')}
\nonumber\\ 
\\
V^{(an)}(l)&=&e^2M_{2ii}^{(an)}\frac{1}{(p_1^++p_2^+)}
(\Delta_{p_1,-p_2}+\Delta_{p_4,-p_3})f_{p_ip_f}\cdot
\int_0^l dl'\frac{f_{p_1,-p_2}(l')f_{p_4,-p_3}(l')}{f_{p_ip_f}(l')}
\nonumber 
\; . \leea{gi13}
This gives rise to 
\bea
V_{gen}^{(ex)}(l) &\hspace{-5em}=\hspace{-5em}& -e^2M_{2ii}^{(ex)}\frac{1}{(p_1^+-p_3^+)} \,
\frac{1}{2} \left( \frac{1}{\Delta_{p_1,p_3}} + \frac{1}{\Delta_{p_4,p_2}} \right) \, \cdot
\left( 1- {\rm e}^{-l\cdot 2\Delta_{p_1,p_3}\Delta_{p_4,p_2}} \right) \, \cdot
{\rm e}^{-l\Delta_{p_ip_f}^2} \nn \\
\\
V_{gen}^{(an)}(l) &\hspace{-5em}=\hspace{-5em}& e^2M_{2ii}^{(an)}\frac{1}{(p_1^++p_2^+)} \,
\frac{1}{2} \left( \frac{1}{\Delta_{p_1,-p_2}}+\frac{1}{\Delta_{p_4,-p_3}} \right) \, \cdot
\left( 1 - {\rm e}^{-l\cdot 2\Delta_{p_1,-p_2}\Delta_{p_4,-p_3}} \right) \, \cdot
{\rm e}^{-l\Delta_{p_ip_f}^2} \nn
\; , \leea{gi14}
where 
\be
\Delta_{p_ip_f}\equiv p_1^-+p_2^--p_3^--p_4^-=
\Delta_{p_1,p_3}-\Delta_{p_4,p_2}
=\Delta_{p_1,-p_2}-\Delta_{p_4,-p_3}
\lee{gi14a}
due to momentum conservation in $'+'$ and 'transversal' directions.

For energy conserving processes, i.e. when
$\Delta_{p_ip_f} = 0$, the unity in \eq{gi14} gives the result of the
standard renormalization procedure. This interaction containes
divergencies in the form of small energy denominator,
as is general for perturbative approach. This problem
is cured in the method of flow equations (and also by
similarity transformations) by the proper choice of the generator
$\eta$ (Appendix C). The divergencies in \eq{gi14} effectively are cancelled
by the exponential factor in the bracket $(1 \!- \!\exp)$ 
as long as the leading order $ee\gamma$ coupling is not completely eliminated 
(i.e. for the finite cutoff $l_{\lambda}$).

We rewrite the renormalized to the second order ${\bf O(e^2)}$
{\bf effective interaction}, \eq{gi14}, as
\bea
V_{gen,\la}^{(ex)} &=& -e^2 N_{1,\la} \,
\frac{1}{2} \left( \frac{1}{\tilde{\De}_1} + \frac{1}{\tilde{\De}_2} \right) \, \cdot
\left( 1 - {\rm e}^{-2 \frac{\De_1}{\la^2} \frac{\De_2}{\la^2}} \right) \, \cdot
{\rm e}^{- \left( \frac{M_0^2-M_0^{'2}}{\la^2} \right)^2} \nn \\
\\
V_{gen,\la}^{(an)} &=& e^2 N_{2,\la} \,
\frac{1}{2} \left( \frac{1}{M_0^2} + \frac{1}{M_0^{'2}} \right) \, \cdot
\left( 1 - {\rm e}^{-2 \frac{M_0^2}{\la^2} \frac{M_0^{'2}}{\la^2}} \right) \, \cdot
{\rm e}^{- \left( \frac{M_0^2-M_0^{'2}}{\la^2} \right) ^2} \nn
\; , \leea{gi15}
where we have introduced
\bea
& P^{+ 2}M_{2ii,\la}^{(ex)} = - N_1 \quad;\qquad 
P^{+ 2}M_{2ii,\la}^{(an)} = N_2 & \nn \\
\nn \\
& \De_{p_1 p_3} = \frac{\De_1}{P^+} = \frac{\widetilde{\De}_1}
{(x' - x) P^+} \quad;\qquad
 \De_{p_4 p_2} = \frac{\De_2}{P^+} = \frac{\widetilde{\De}_2}
{(x' - x) P^+}; & \nn \\
\nn \\
& \De_{p_1,-p_2} = \frac{M^2_0}{P^+} \quad;\qquad
 \De_{p_4,-p_3} = \frac{{M'}^2_0}{P^+} & \nn \\
\leea{gi16}
(see Appendix B for the explicit definition of these quantities
in the light-front frame).

The expression \eq{gi15} is written for the rescaled value
of the potential $V\rightarrow P^{+ 2}V$,
and the cutoff is defined in units of the total momentum $P^+$,
i.e. $\la\rightarrow\frac{\la^2}{P^+}$, with $l=1/\la^2$.
The spin structure of the interaction is carried by
the matrix elements $M_{2ii}$, defined in Appendix B.

\begin{figure}
\setlength{\unitlength}{1mm}
\begin{picture}(170,71)
\put(19,36){\makebox(56,34.61){ \loadeps{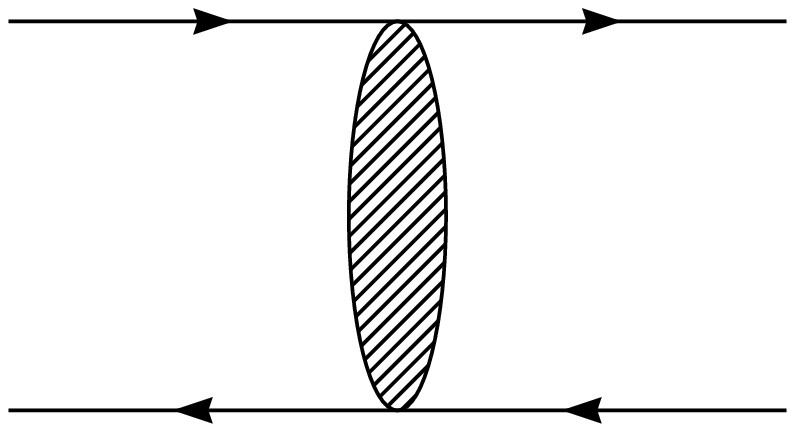} }}
\put(18,67){$p_1\;(x,k^\bot)$}
\put(50,67){$p_3\;(x',k'^\bot)$}
\put(18,38){$p_2\;(1\!-\!x,-k^\bot)$}
\put(50,38){$p_4\;(1\!-\!x',-k'^\bot)$}
  \put(75,36){\makebox(56,34.61){ \loadeps{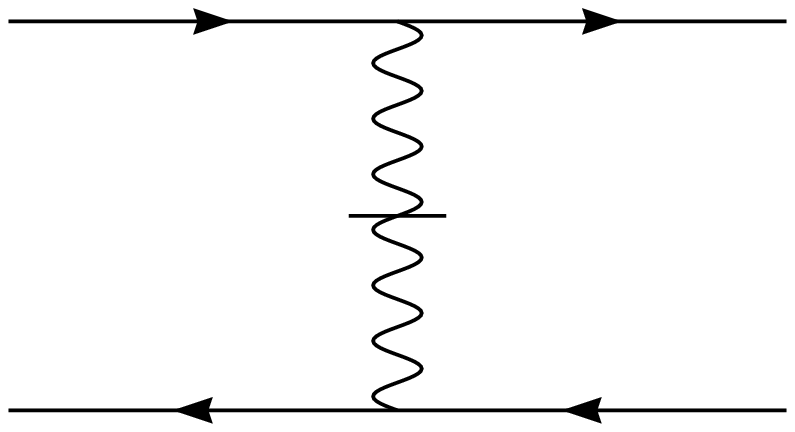} }}
\put(73,53.305){$+$}
\put(38,0){\makebox(56,34.61){ \loadeps{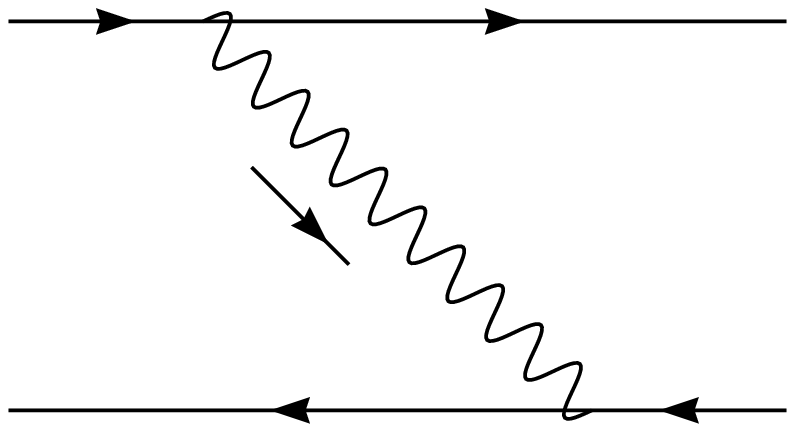} }}
  \put(94,0){\makebox(56,34.61){ \loadeps{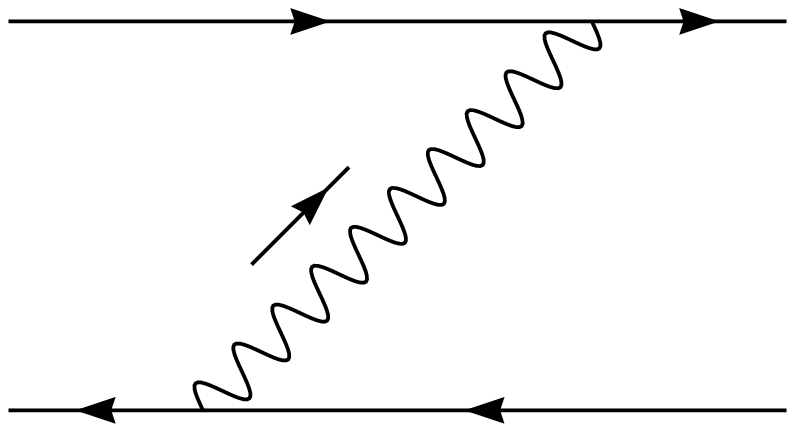} }}
\put(36,17.305){$+$}
\put(92,17.305){$+$}
\end{picture}
\caption{The renormalized to the second order electron-positron
interaction in the exchange channel; diagrams correspond to generated, 
instantaneous interactions and two perturbative photon exchanges 
with respect to different time ordering.}
\label{reneebarint}
\end{figure}

We summarize the {\bf instantaneous interaction} in both channels
to the order ${\bf O(e^2)}$, cf. \fig{feynrules},
\bea
&& V_{inst}^{(ex)} = -\frac{4e^2}{(p_1^+-p_3^+)^2} \;
\delta_{s_1s_3} \delta_{s_2s_4} \; \exp(-l\Delta_{p_ip_f}^2) \nn \\
\\
&& V_{inst}^{(ex)} = \frac{4e^2}{(p_1^++p_2^+)^2} \;
\delta_{s_1\bar{s}_2} \delta_{s_3\bar{s}_4} \; \exp(-l\Delta_{p_ip_f}^2)
\nonumber
\; , \leea{gi17}
where we have used
\mbox{$\chi_{s_3}^+ \chi_{\bar{s}_2}^+ \bbbone \chi_{s_1} \chi_{\bar{s}_4} =
\delta_{s_1 s_3} \delta_{s_2 s_4} + \delta_{s_1 \bar{s}_2} \delta_{s_3 \bar{s}_4}$}.
For the rescaled potential in the light-front frame (\fig{reneebarint} and Appendix B
\eq{b14} and following) we thus have
\bea
&& V_{inst,\la}^{(ex)} = -\frac{4e^2}{(x-x')^2} \;
\delta_{s_1s_3} \delta_{s_2s_4} \;
\exp\left\{ -\left( \frac{M_0^2-{M'_0}^2}{\la^2} \right)^2 \right\} \nn \\
\\
&& V_{inst,\la}^{(an)} = 4e^2 \;
\delta_{s_1\bar{s}_2} \delta_{s_3\bar{s}_4} \;
\exp\left\{ -\left( \frac{M_0^2 - {M'_0}^2}{\la^2} \right)^2 \right\} \nn
\; , \leea{gi18}
where the notations of eqs{gi14a}{gi16} are implied.

\subsubsection{Renormalization issues}

As was discussed above after \eq{fe13} the commutator $[\eta^{(1)},H_{ee\ga}]$ also
contributes to the self-energy term, giving rise to the renormalization
of fermion and photon masses to the second order.
The flow equation for the electron (photon) light-cone energy $p^-$ is
\be
\frac{dp^-}{dl}=<[\eta^{(1)},H_{ee\ga}]>_{self energy}
\; , \lee{ri1}
where the matrix element is calculated between the dressed single
electron (photon) states $<p',s'|...|p,s>$. We drop the finite part
and define $\de p_{\la}^- = p^-(l_{\la})-<|H_0|>$. Integration over the
finite range gives
\be
\de p_{\la}^--\de p_{\La}^-=\int_{l_{\La}}^{l_{\la}}\\
<[\eta^{(1)},H_{ee\ga}]>_{self energy}dl'\\
=-\frac{(\de\Sigma_{\la}(p)-\de\Sigma_{\La}(p))}{p^+}
\; , \lee{ri2}
that defines the cutoff dependent self energy $\de\Sigma_{\la}(p)$.
The mass correction and wave function renormalization constant
correspondingly are given, cf. \cite{ZhHa}, as
\bea
\de m_{\la}^2 &=& \left. p^+\de p^- \right|_{p^2=m^2}
 =-\de\Si_{\la}(m^2) \nn \\
Z_2 &=& \left. 1 + \frac{\partial \de p^-}{\partial p^-} \right|_{p^2=m^2}
\; . \leea{ri3}
The on-mass-shell condition is defined through the mass $m$ in the
free Hamiltonian $H_0$.

We show further, that to the second order $O(e^2)$ the electron and photon
masses and corresponding wave function renormalization constants
in the renormalized Hamiltonian vary in accordance with the result of $1$-loop
renormalization group equations. This can serve as evidence for the 
equivalence of the herediscussed method of flow equations and Wilson's
renormalization scheme. Therefore we have rewritten
the mass correction $\de m_{\la}^2$ through the self energy term,
arising in $1$-loop calculations of ordinary perturbative theory. The negative
overall sign stems from our definition of the flow parameter,
namely for $\De l>0$ we are lowering the cutoff 
\mbox{$dl=-\frac{2}{\la^3}d\la$}. 
 
We start with the bare cutoff mass \mbox{$m_{\La}^2=m^2+\de M_{\La}^{(2)}$},
where \mbox{$\de M_{\La}^{(2)}$} is the second order mass counterterm.
According to \eq{ri2} the electron (photon) mass runs
\be
m_{\la}^2 = m_{\La}^2 - [\de\Sigma_{\la}(m^2) - \de\Sigma_{\La}(m^2)]
\lee{ri4}
defining, due to renormalizability, the counterterm
\mbox{$\de M_{\La}^{(2)} =\de m_{\La}^2 = -\de\Sigma_{\La}(m^2)$}
and the dependence of the renormalized mass on the cutoff $\la$
\be
m_{\la}^2 = m^2 + \de m_{\la}^2 = m^2 - \de\Sigma_{\la}
\; . \lee{ri5}
We calculate explicitly the self-energy term.
The {\bf electron} energy correction contains several terms
\be
\de p_{\la}^-
= <p', s'|H - H_0|p, s>
= \left(\sum_{n=1}^3 \de p_{\la n}^-\right) \cdot \de^{(3)}(p-p') \de_{s s'}
\; . \lee{ri6}
The first term comes from the commutator $[\eta{(1)}, H_{ee\ga}]$
\be
\de p_{1\la}^-=-\int_{l_{\la}}^{\infty}\\
<[\eta^{(1)},H_{ee\ga}]>_{self energy}dl'=-\frac{\de\Sigma_{1\la}(p)}{p^+}
\; ; \lee{ri7}
it reads, cf. \eq{d9} in Appendix D,
\bea
\de p_{1\la}^- &=& e^2 \int \frac{d^2k^{\bot} dk^+}{2(2\pi)^3}
\frac{\th (k^+)}{k^+} \th (p^+-k^+) \nn \\
&& \times \Ga^i(p - k, p, -k) \Ga^i(p, p - k, k) \,
\frac{1}{p^- - k^- - (p-k)^-} \times (-R)  
\; . \leea{ri8}
This term explicitly depends on the cutoff $\la$ through the
$f$-function, that plays the role of a regulator in the loop integration
\be
R_{\la}=f_{p,k,\la}^2=\exp\left\{-2\left(\frac{\De_{p,k}}{\la}\right)^2\right\}
\; . \lee{ri9}
\Eq{ri8} corresponds to the first diagram in \fig{eselfen}.

Two instantaneous diagrams, the second and third in \fig{eselfen}, contribute 
cutoff independent (constant) terms. Formally one can write
\be
\frac{dp^-}{dl} \; = \; <[\eta^{(2)},H_0]>|_{self energy} \;
 = \; <\hat{O}\hat{O}^+> \, \frac{dV_{pp'}^{inst}}{dl}
\; , \lee{ri10}
where \mbox{$<\hat{O}\hat{O}^+>$} stands for both the fermion
and boson contraction (i.e. \mbox{$<b_p b_p^+> = \th(p^+)$} and
\mbox{$<a_k a_k^+> = \th(k^+)$}, respectively); 
and \mbox{$V_{pp'}^{inst}(l) = f_{pp'}(l) \, V^{inst}(l=0)$} is defined in ().
This gives rise to
\be
\de p_{\la n} \; = \; 
 <\hat{O}\hat{O}^+> \cdot V_{pp}^{inst}(l) \; = \;
 <\hat{O}\hat{O}^+> \cdot V^{inst}(l=0)
\lee{ri11}
for $n=2,3$. This means that $\de p_{\la n}$ defines together with
$\de p_{\la 1}(l\!=\!0)$ the initial condition for the total energy
correction, \eq{ri6}.

In perturbative theory the instantaneous diagrams arise from
normal-ordering Hamiltonian at $l=0$, and, in principle, must accompany
the first diagram for any $l$. In what follows we use for 
the instantaneous terms the same regulator $R$, \eq{ri9},
\bea
\de p_{2\la}^- &=& e^2\int \frac{d^2k^{\bot}dk^+}{2(2\pi)^3} \,
\frac{\th (k^+)}{k^+} \, \frac{\si^i\si^i}{[p^+-k^+]} \times (-R) \nn \\
\de p_{3\la}^- &=& e^2\int \frac{d^2k^{\bot}dk^+}{2(2\pi)^3} \, \th (k^+) \,
\frac{1}{2} \left( \frac{1}{[p^+ - k^+]^2} - \frac{1}{[p^+ + k^+]^2} \right) \times (-R)
\; . \leea{ri12}
We define the set of coordinates
\bea
x &=& \frac{k^+}{p^+} \nn \\
k &=& (xp^+, xp^{\bot} + \kappa^{\bot})
\; , \leea{ri13}
where \mbox{$p = (p^+, p^{\bot})$} is the external electron momentum.
Then the electron self energy diagrams, \fig{eselfen}, cf. also \eq{d13}
in Appendix D, contribute
\bea
\hspace{-10mm} p^+\de p_{1\la}^- &=& -\frac{e^2}{8\pi^2}
 \int_0^1 dx \int d\kappa_{\bot}^2 \nn \\
&& \hspace{10mm} \times \left[
 \frac{p^2 - m^2}{\kappa_{\bot}^2 + f(x)} \left( \frac{2}{[x]} - 2 + x \right)
 -\frac{2m^2}{\kappa_{\bot}^2 + f(x)} + \left( \frac{2}{[x]^2} + \frac{1}{[1-x]} \right)
 \right]
 \times (-R) \nn \\
f(x) &=& xm^2 - x(1-x) p^2
\leea{ri14}
and
\bea
p^+ \de p_{2\la}^- &=& \frac{e^2}{8\pi^2} \int_0^{\infty}dx \int d\kappa_{\bot}^2
\left( \frac{1}{[x][1-x]} \right) \times (-R) \nn \\
&& \hspace{-7mm} \rightarrow \; \frac{e^2}{8\pi^2} \int_0^1dx \int d\kappa_{\bot}^2
\left( \frac{1}{[x]} \right) \times (-R) \nn \\
p^+ \de p_{3\la}^- &=& \frac{e^2}{8\pi^2} \int_0^{\infty}dx \int d\kappa_{\bot}^2
\left( \frac{1}{[1-x]^2} - \frac{1}{(1+x)^2} \right) \times (-R) \nn \\
&& \hspace{-7mm} \rightarrow \; \frac{e^2}{8\pi^2} \int_0^1dx \int d\kappa_{\bot}^2
\left( \frac{2}{[x]^2} \right) \times (-R)
\; ; \leea{ri15}
for details we refer to Appendix D. 
Note, that the transformation in the integrals over $x$
is performed {\it before} the regulator is taken into account \cite{ZhHa}.
(In the second integral the electron momentum is replaced by the gluon
one due to momentum conservation). The brackets '\mbox{\boldmath{$[\;]$}}'
denote the principle value prescription, defined later in \eq{ri21}.

The loop integral over $k$ \eqs{ri14}{ri15}
contains two types of divergencies: UV in the transversal
coordinate $\kappa^{\bot}$ and IR in the longitudinal component $k^+$.
The physical value of mass must be IR-finite.
We show, that the three relevant diagrams together 
in fact give an IR-finite value for the renormalized mass; this enables to
determine {\it counterterms independent of longitudinal momentum}. 
In the wave function renormalization constant, however, the IR-singularity
is still present.
    
Define 
\be
\de_1=\frac{p^+}{P^+}
\; , \lee{ri16}
where $P=(P^+,P^{\bot})$ is the positronium momentum, $p$ the electron momentum. 
The transversal UV divergency is regularized through the unitary
transformation done, i.e. by the regulator $R$, \eq{ri9}
\be
R_{\la} =
\exp\left\{ -\left( \frac{\tilde{\De}_{p,k}}{\la^2\de_1} \right)^2 \right\}
\; \approx \;
\th(\la^2 \de_1 - |\tilde{\De}_{p,k}|)
\; , \lee{ri17}
where the cutoff is rescaled and defined in units of the
positronium momentum $P^+$, namely \mbox{$\la\rightarrow \sqrt{2}\la^2/P^+$},
and \mbox{$\De_{p,k}=p^--k^--(p-k)^-=\tilde{\De}_{p,k}/p^+$}. The rude approximation
for the exponential through a $\th$-function changes the numerical coefficient
within a few percent; nevertheless it is useful to estimate the integrals
in \eqs{ri14}{ri15} in this way {\it analitically}.
From \eq{ri17} we have for the sum of intermediate (electron and photon) state momenta 
(the external electron is on-mass-shell \mbox{$p^2=m^2$}) 
\be
\frac{\kappa^{\bot 2}}{[x]}+\frac{\kappa^{\bot 2}+m^2}{[1-x]} \; \leq \;
\la^2\de_1+m^2
\lee{ri18}
giving for the regulator
\bea
R_{\la} &=& \th(\kappa^{\bot 2}_{\la max}-\kappa^{\bot 2}) \,
\th(\kappa^{\bot 2}_{\la max})\nn\\
\kappa^{\bot 2}_{\la max}&=&x(1-x)\la^2\de_1-x^2m^2
\leea{ri19}
and \mbox{$\th(\kappa^{\bot 2}_{\la max})$} leads
to the additional condition for the longitudinal momentum
\bea
&& 0\leq x\leq x_{max} \nn \\
&& x_{max}=\frac{1}{1+m^2/(\la^2 \de_1)}
\leea{ri20}
implying that the singularity of the photon longitudinal momentum
for $x\rightarrow 1$ is regularized by the function $R_{\la}$.
This is the case due to the nonzero fermion mass present in \eq{ri18}
for the intermediate state with $(1-x)$ longitudinal momentum.
The IR-singularity when $x\rightarrow 0$ is still present; it
is treated by the principle value prescription \cite{ZhHa}
\be
\frac{1}{k^+}=\frac{1}{2}\left( \frac{1}{k^++i\varep P^+}
+\frac{1}{k^+-i\varep P^+}\right)
\; , \lee{ri21}
where $\varep=0_+$, and $P^+$ is the longitudinal part of the positronium
momentum (used here as typical momentum in the problem being under
discussion). This defines the bracket '\mbox{\boldmath{$[\;]$}}'
in \eqs{ri14}{ri15}
\be
\frac{1}{[x]}=\frac{1}{2} \left( \frac{1}{x+i\frac{\varep}{\de_1}}+\\
\frac{1}{x-i\frac{\varep}{\de_1}} \right)
\; . \lee{ri22}
Making use of both regularizations for 
transversal and longitudinal components, we have for the first diagram,
\eq{ri14},
\bea
\de m_{1\la}^2 &\hspace{-1em}=\hspace{-1em}& p^+ \de p^- |_{p^2=m^2} \nn \\
\de m_{1\la}^2 &\hspace{-1em}=\hspace{-1em}&-\frac{e^2}{8\pi^2}
\left\{ 3m^2 \ln \left( \frac{\la^2\de_1+m^2}{m^2} \right)
+ \frac{\la^2 \de_1}{\la^2\de_1+m^2} \left( \frac{3}{2}\la^2\de_1+m^2 \right)
-2 \la^2 \de_1 \ln \left( \frac{\la^2\de_1}{\la^2\de_1+m^2} \frac{\de_1}{\varep} \right)
\right\} \nn \\
\leea{ri23}
Note, that the third term has the mixing UV and IR divergencies.
Combining the three relevant diagrams, \fig{eselfen}, and integrating
with the common regulator, one obtains for the {\bf electron mass correction} 
\bea
\de m_{\la}^2 &=& p^+(\de p_1 + \de p_2 + \de p_3)|_{p^2 = m^2} =
-\de \Sigma_{\la}(m^2) \nn \\
\de m_{\la}^2 &=& -\frac{e^2}{8\pi^2}
\left\{ 3m^2 \ln \left( \frac{\la^2\de_1+m^2}{m^2} \right)
-\frac{\la^2 \de_1 m^2}{\la^2 \de_1 + m^2} \right\}
\; . \leea{ri24}
The mass correction is IR-finite (that gives rise to IR-finite
counterterms) and contains only a logarithmic UV-divergency. Namely,
when \mbox{$\la\de_1 \rightarrow\La \gg m$}
\be
\de m_{\La}^2=-\frac{3e^2}{8\pi^2}m^2 \ln \frac{\La^2}{m^2}
\; . \lee{ri25}
It is remarkable that we reproduce with the cutoff condition of \eq{ri18}
the standard result of covariant perturbative theory calculations  
including its global factor $3/8$. As was mentioned above, the difference
in sign, as compared with the $1$-loop renormalization group result, 
comes from scaling down from high to low energies 
in the method of flow equations.     

The similar regularization for the intermediate state momenta
in the self-energy integrals, called global cutoff scheme, 
was introduced by W.~M.~Zhang and A.~Harindranath \cite{ZhHa}.
In our approach the UV-regularization, that defines the concrete form of 
the regulator $R$, arises naturally from the method of flow equations, 
namely from the unitary transformation performed, where the generator
of the transformation is chosen as the commutator $\eta=[H_0,H]$.
Note also, that the regulator $R$, \eq{ri17}, in general
is independent of the electron momentum $p^+$ (rescaled cutoff
\mbox{$\la\de_1\longrightarrow\la$}), and therefore is boost invariant.

For the wave function renormalization constant, \eq{ri3}, one has
\be
\left. \frac{\partial\de p^-}{\partial p^-} \right|_{p^2=m^2} =
-\frac{e^2}{8\pi^2}\int_0^1\int d\kappa_{\bot}^2 \left[
\frac{2\frac{1}{x}-2+x}{\kappa_{\bot}^2+f(x)}
-\frac{x(1-x)2m^2}{(\kappa_{\bot}^2+f(x))^2}\right]_{p^2=m^2}
\times(-R)
\; , \lee{ri26}
that together with the regulator $R$, \eq{ri17}, results
\bea
Z_2 &\hspace{-0.5em}=\hspace{-0.5em}& 1 - \frac{e^2}{8\pi^2} \left\{
\ln\frac{\la^2\de_1}{m^2} \cdot \left( \frac{3}{2}-2 \ln\frac{\de_1}{\varep} \right)
+ \ln\frac{\de_1}{\varep} \cdot \left( 2 - \ln\frac{\de_1}{\varep} \right)
+ F \left( \ln\frac{\la^2\de_1}{\la^2\de_1+m^2}; \frac{\la^2\de_1}{\la^2\de_1+m^2} \right)
\right\} \nn \\
F &\hspace{-0.5em}=\hspace{-0.5em}& \ln\frac{\la^2\de_1}{\la^2 \de_1+m^2} \left(
\frac{1}{2} - \ln\frac{\la^2 \de_1}{\la^2 \de_1+m^2} \right)
+\frac{1}{2} \frac{\la^2 \de_1}{\la^2 \de_1 + m^2}
- 2 + 2 \int_0^{x_{max}}dx \frac{\ln x}{x-1}
\; . \leea{ri27}
As \mbox{$\la\de_1 \rightarrow \La \gg m$} the function $F$ tends to a constant
\be
F|_{\La \gg m} = C = -\frac{3}{2} +\frac{\pi^2}{3}
\; . \lee{ri28}
Therefore, by dropping the finite part, we obtain
\bea
Z_2 &=& 1 - \frac{e^2}{8\pi^2}
\left\{\ln\frac{\La^2}{m^2} \cdot \left( \frac{3}{2}
-2 \ln\frac{1}{\varep} \right)
+ \ln\frac{1}{\varep} \left( 2-ln\frac{1}{\varep} \right) \right\}
\; , \leea{ri29} 
where we have rescaled 
\mbox{$\frac{\varepsilon}{\de_1}\rightarrow\varepsilon$}. 
The electron wave function renormalization constant contains
logarithmic UV and IR divergencies {\it mixed}, together with {\it pure}
logarothmic IR divergencies. 
We mention, that the value of $Z_2$ is not sensitive to what kind of regulator
is applied; the same result for $Z_2$ was obtained with another choice of
regulator \cite{ZhHa}.

We proceed with renormalization to the second order in the photon sector.
The diagrams that contribute to the photon self energy are shown in \fig{photselfen}.
The commutator \mbox{$[\eta^{(1)},H_{ee\ga}]$}, corresponding to the first diagram,
gives rise to (cf. \eq{d21} in Appendix D)
\bea
\de q_{1\la}^ - \, \de^{ij} &=&
\frac{1}{[q^+]} e^2 \int \frac{d^2k^{\bot}dk^+}{2(2\pi)^3} \,
\th(k^+) \th(q^+-k^+) \\
&&\hspace{3em} \times \, Tr\left[\Ga^i(k,k-q,q) \Ga^j(k-q,k,-q)\right] \,
\frac{1}{q^- - k^- - (q-k)^-} \times(-R) \nn
\; , \leea{ri30}
where momenta are given in \fig{photselfen} , and the regulator is
\be
R_{\la}=f_{q,k,\la}^2=\exp\left\{-2 \left(\frac{\De_{q,k}}{\la} \right)^2\right\}
\; . \lee{ri31}
In full analogy with the electron self energy this also defines the regulator
for the second diagram with the instantaneous interaction, see \fig{photselfen},
\be
\de q_{2\la}^- \, \de^{ij} = \frac{1}{[q^+]} e^2 \int \frac{d^2k^{\bot}dk^+}{2(2\pi)^3}
\th(k^+) \, Tr(\si^i \si^j)
\left( \frac{1}{[q^+-k^+]} - \frac{1}{[q^++k^+]} \right) \times(-R)
\; . \lee{ri32}
We define the set of coordinates
\bea
\frac{(q-k)^+}{q^+} &=& x  \nn \\
k &=& ((1-x)q^+, (1-x)q^{\bot}+\kappa^{\bot}) \nn \\
(q - k) &=& (xq^+, xq^{\bot} - \kappa^{\bot})
\; , \leea{ri33}
where $q = (q^+, q^{\bot})$ is the external photon momentum. Then two diagrams
contribute (for details see Appendix D, \eq{d25}):
\bea
q^+ \, \de q_1^- &=&
-\frac{e^2}{8\pi^2} \int_0^1dx \int d\kappa_{\bot}^2 \nn \\
&& \times \left\{
 \frac{q^2}{\kappa_{\bot}^2+f(x)} \left( 2x^2 - 2x + 1 \right) +
 \frac{2m^2}{\kappa_{\bot}^2+f(x)}
 + \left( -2 + \frac{1}{[x][1-x]} \right)
\right\} \times (-R) \nn \\
f(x) &=& m^2 - x(1-x)q^2 \nn \\
q^+ \, \de q_2^- &=& \frac{e^2}{8\pi^2} \int_0^\infty dx \int d\kappa_{\bot}^2
\left( \frac{1}{[1-x]}-\frac{1}{1+x} \right) \times (-R) \nn \\
&&\hspace{-1.5em} \rightarrow \;
-\frac{e^2}{8\pi^2} \int_0^1dx \int d\kappa_{\bot}^2 \frac{2}{[x]} \times(-R)
\; . \leea{ri34}
Note, that the transformation in the second integral is done {\it before}
the regularization (by regulator the $R$) is performed \cite{ZhHa}.

Making use of the same approximation for the regulator as in the electron
sector, we obtain for the sum of intermediate (two electron) state momenta
\bea
&& \frac{\kappa_{\bot}^2+m^2}{x} + \frac{\kappa_{\bot}^2+m^2}{1-x}
 \leq \la^2\de_2 \nn \\
&& \de_2 = \frac{q^+}{P^+}
\; , \leea{ri35}
where the photon is put on mass-shell $q^2=0$ and the rescaled
cutoff \mbox{$\la \rightarrow \sqrt{2} \la^2/P^+$} has been used.
This condition means for the transversal integration
\bea
R_{\la} &=& \th(\kappa_{\la max}^{\bot 2}-\kappa^{\bot 2}) \,
\th(\kappa_{\la max}^{\bot 2}) \nn \\
\kappa_{\la max}^{\bot 2} &=& x(1-x)\la^2 \de_2-m^2
\leea{ri36} 
and for the longitudinal integration
\bea
&& x_1 \leq x \leq x_2 \nn \\
&& x_1 = \frac{1-r}{2} \approx \frac{m^2}{\la^2\de_2} \nn \\
&& x_2 = \frac{1+r}{2} \approx 1 - \frac{m^2}{\la^2\de_2} \nn \\
&& r = \sqrt{1 - \frac{4m^2}{\la^2\de_2} }
\; , \leea{ri37}
where the approximate value is when $m \ll \la$.
This shows that the condition of \eq{ri35} for two electrons with masses $m$ 
removes the light-front infrared singularities from \mbox{$x \rightarrow 0$}
and \mbox{$x\rightarrow 1$}. Thus, both UV and IR divergencies are regularized
by the regulator R, \eq{ri36}.

The mass correction arising from the first diagram, \eq{ri34}, is
\be
\de m_{1\la}^2 = \frac{e^2}{8\pi^2} \, \frac{2}{3} \, \la^2 \de_2
\left( 1-\frac{4m^2}{\la^2\de_2} \right)^{3/2}
\; . \lee{ri38}
Combining together both diagrams with the same regulator, \eq{ri34}, we obtain
\be
\de m_{\la}^2 = \frac{e^2}{8\pi^2} \, \left( \frac{5}{3} \la^2 \de_2 \, r
- \frac{8}{3} m^2 \, r - 2m^2 \, \ln\frac{1+r}{1-r} \right)
\; , \lee{ri39}
where $r$ is defined in \eq{ri37}.
The result shows that the mass correction involves the quadratic
and logarithmic UV divergencies, i.e. as $\la\de_2\rightarrow\La\gg m$
\be
\de m_{\La}^2 = \frac{e^2}{8\pi^2} \left( \frac{5}{3} \La^2 
-2m^2 \, \ln\frac{\La^2}{m^2} \right)
\; . \lee{ri40}
The wave function renormalization constant is defined through 
\be
\left. \frac{\partial\de q^-}{\partial q^-} \right|_{q^2=0} = 
- \frac{e^2}{8\pi^2} \int_0^1dx \int d\kappa_{\bot}^2
\left. \left\{ \frac{2x^2-2x+1}{\kappa_{\bot}^2+f(x)}
+ \frac{2m^2x(1-x)}{(\kappa_{\bot}^2+f(x))^2}
\right\} \right|_{q^2=0} \times (-R)
\; , \lee{ri41}
that, with the regulator $R$, \eq{ri36}, results
\be
Z_2 = 1 - \frac{e^2}{8\pi^2} \left( -\frac{2}{3}\ln\frac{1+r}{1-r}
+ \frac{10}{9} \, r + \frac{8}{9} \frac{m^2}{\la^2\de_2}\, r \right)
\; . \lee{ri42}
The photon wave function renormalization constant contains only
logarithmic UV divergency, indeed as \mbox{$\la\de_2\rightarrow\La\gg m$} 
\bea
&& Z_2=1-\frac{e^2}{8\pi^2}(-\frac{2}{3}\ln\frac{\La^2}{m^2})
\leea{ri43}
and is free of IR divergencies (as is expected from the form of the
regulator $R$, \eq{ri35}).

\subsection{Renormalized theory to the order \boldmath$e^2$
 (at the scale \boldmath$\la$) and Feynmann rules}

\begin{figure}
\setlength{\unitlength}{1mm}
\begin{picture}(170,219)
\put(0,185){\makebox(56,34.61){ \loadeps{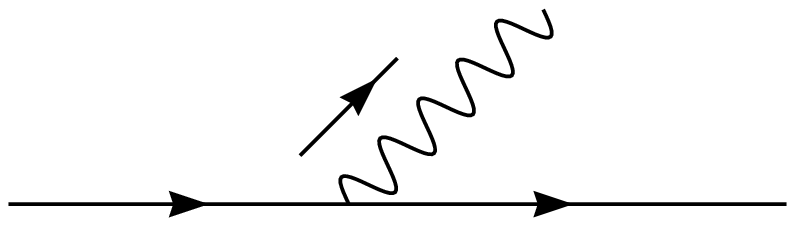} }}
  \diagform{180}{
    \bea
    &&\hspace{-15mm} -e_{\la}\,f_{p_i p_f, \la} \cdot \ch^+_2
       \Ga^i_\la(p_1,p_2,k) \ch_1 \varep^{i \ast} \nn \\
    && \nn \\
    &&\hspace{-5mm} \Ga^i_\la(p_1,p_2,k) =   
         2 \frac{k^i}{[k^+]} - \frac{\si \cdot p_2^\bot - im_{\la}}{[p_2^+]} \si^i
                         - \si^i \frac{\si \cdot p_1^\bot + im_{\la}}{[p_1^+]} \nn \\
    &&\hspace{-3mm} i=1,2 \nn                           
    \eea 
    }
\put(10,193){$p_1$}
\put(42,193){$p_2$}
\put(34,210){$k \; (i)$}
\put(0,140){\makebox(56,34.61){ \loadeps{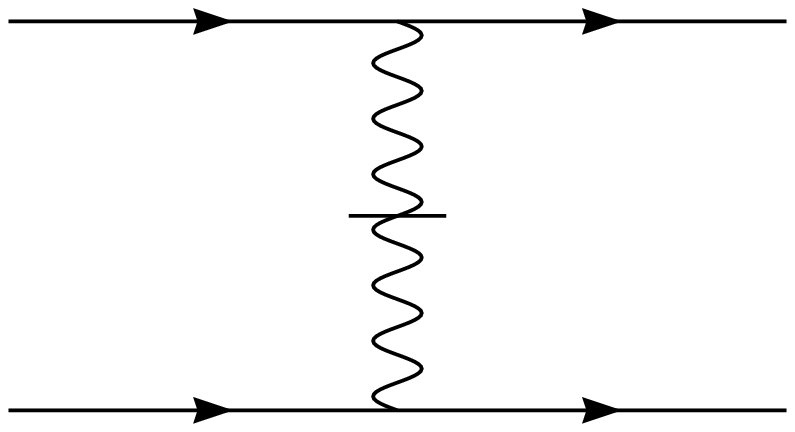} }}
  \diagform{135}{
    \[
    \hspace{-55mm} e_{\la}^2\,f_{p_i p_f, \la} \cdot \ch^+_3 \ch^+_4
       \frac{4}{[p_1^+ - p_2^+]^2} \ch_1 \ch_2
    \] 
    }
\put(10,171){$p_1$}
\put(42,171){$p_2$}
\put(10,142){$p_3$}
\put(42,142){$p_4$}
\put(0,95){\makebox(56,34.61)[r]{ \loadeps{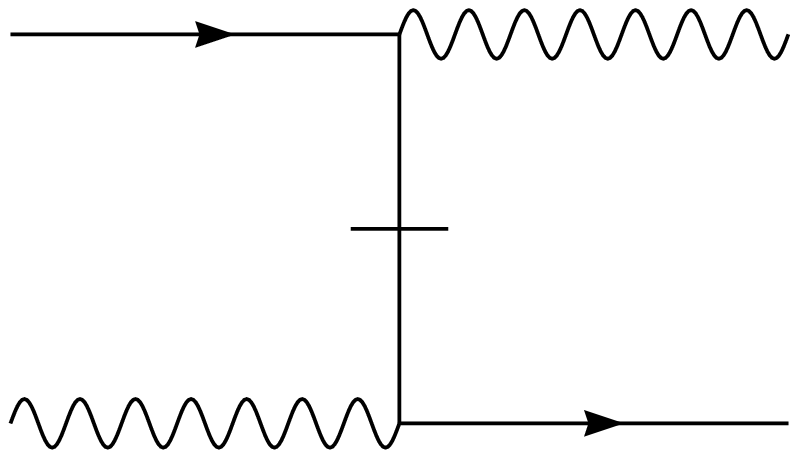} }}
  \diagform{90}{
    \[
    \hspace{-55mm} e_{\la}^2\,f_{p_i p_f, \la} \cdot \ch^+_2
       \frac{\si^j \si^i}{[p_1^+ - k_1^+]} \ch_1 \varep^{i \ast} \varep^j
    \] 
    }
\put(10,126){$p_1$}
\put(42,126){$k_1 \; (i)$}
\put(10,97){$k_2 \; (j)$}
\put(42,97){$p_2$}
\put(0,50){\makebox(56,34.61){ \loadeps{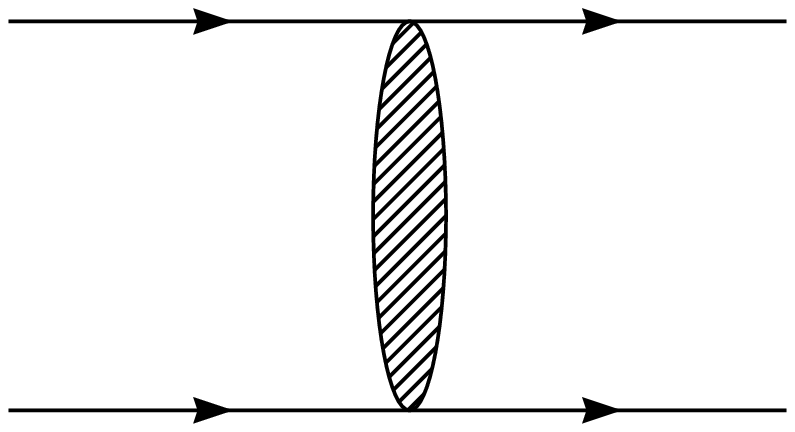} }}
  \diagform{45}{
    \bea
    &&\hspace{-17mm} e_{\la}^2\,f_{p_i p_f, \la} \cdot M_{2ij,\la} \, \de^{ij} \cdot
      \frac{1}{[p_1^+ - p_2^+]} \nn \\
    &&\hspace{-12mm} 
      \times \frac{1}{2} \left( \frac{1}{\De_{p_1p_2\la}} + \frac{1}{\De_{p_4p_3\la}} \right) \,
      \left( 1 - \exp{\left\{ -2 \, \frac{\De_{p_1p_2\la} \cdot \De_{p_4p_3\la}}{\la^2} \right\} }
       \right) \nn \\
    && \nn \\
    &&\hspace{-7mm} M_{2ij,\la} = \biggl(\ch^+_2 \Ga^i_\la(p_1,p_2,p_1 \!-\! p_2)\ch_1\biggr) \nn \\
    &&\hspace{13mm} \times \biggl(\ch^+_4 \Ga^j_\la(p_3,p_4,-(p_1 \!-\! p_2)) \ch_3\biggr) \nn
    \eea 
    }
\put(10,81){$p_1$}
\put(42,81){$p_2$}
\put(10,52){$p_3$}
\put(42,52){$p_4$}
\put(0,5){\makebox(56,34.61){ \loadeps{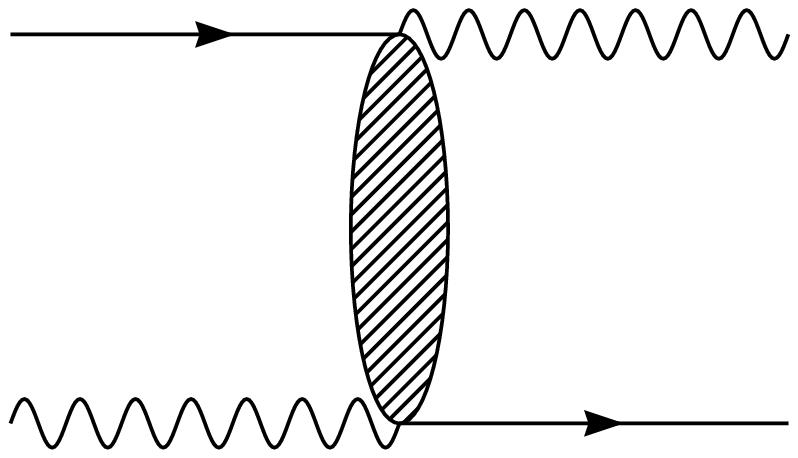} }}
  \diagform{0}{
    \bea
    &&\hspace{-17mm} e_{\la}^2\,f_{p_i p_f, \la} \cdot \widetilde{M}_{2ij,\la} \,
      \varep^{i \ast} \varep^j \nn \\
    &&\hspace{-12mm} 
      \times \frac{1}{2} \left( \frac{1}{\De_{p_1k_1\la}} + \frac{1}{\De_{p_2k_2\la}} \right) \,
      \left( 1 - \exp{\left\{ -2 \, \frac{\De_{p_1k_1\la} \cdot \De_{p_2k_2\la}}{\la^2} \right\} }
       \right) \nn \\
    && \nn \\
    &&\hspace{-7mm} \widetilde{M}_{2ij,\la} = 
       \ch^+_2 \Ga^i_\la(p_1,p_1 \!-\! k_1,k_1) \: \Ga^j_\la(p_1 \!-\! k_1,p_2,k_2) \ch_1 \nn
    \eea 
    }
\put(10,36){$p_1$}
\put(42,36){$k_1 \; (i)$}
\put(10,7){$k_2 \; (j)$}
\put(42,7){$p_2$}
\end{picture}
\caption{ Renormalized to the second order $O(e^2)$ light cone theory  
(the UV cutoff is $\lambda$).
The photon momenta are $x^+$-ordered, from left to right. The similarity
function \mbox{$f_{p_ip_f,\la} \!=\! \exp(-\De_{p_ip_f\la}^2 \!/\!\la^2)$}
playes the role of regulator,
where \mbox{$\De_{p_ip_f\la} \!=\! \Si p^-_i \!-\! \Si p^-_f$}
(the index \mbox{$`i`$} denotes initial and \mbox{$`f`$} final states)
and \mbox{$\De_{p_1p_2\la} \!=\! p_1^- \!-\! p_2^- \!-\! (p_1 \!-\! p_2)^-$},
\mbox{$p^-\!=\! (p_{\bot}^2 \!+\! m_{\la}^2)\!/\!p^+$}. } 
\label{feynrules}
\end{figure}

We have completed the renormalization of light front QED (LFQED)
to the second order, what enables us to formulate the diagrammatic rules
for the perturbative expansions in $e_0$. We formulate
the Feynman rules (in light front frame) for the renormalized
$x^+$-ordered LF Hamiltonian.
All matrix elements of the renormalized  Hamiltonian $H_r$, namely
the interacting part of $H_r$ with the corresponding LFQED vertices,
are listed in \fig{feynrules} together with their diagrams.
The diagrammatic rules are obtained by direct calculation of matrix
elements between free particle states.

The two last diagrams in \fig{feynrules} correspond to the generated interactions,
arising from the renormalized Hamiltonian considered in different sectors.
Note, that the existence of the $|ee\ga>$-vertex in the renormalized Hamiltonian
at any finite cutoff $\la$ prevents the generated interaction 
to have small energy denominators $\frac{1}{\De_1}$, $\frac{1}{\De_2}$
(the $|ee\ga>$-vertex appears through the exponential factor in the generated
interaction). We mention, that the problem of small energy denominators in
generated terms also is solved in the similarity scheme of Gla\-zek
and Wil\-son \cite{GlWi}. 
 
The two-component LF theory, introduced by Zhang and Harindranath \cite{ZhHa},
as compared with the four-component formalism of Lepage and Brodsky
is formulated purely in terms of physical degrees of freedom,
so that each term corresponds to a real dynamical process. This means,
for instance, that the first of the two generated term diagrams 
must be taken into account in low energy $ee$-scattering,
and the second one describes Compton scattering at low energies.
Further, we use the instantaneous and generated interactions,
namely the second and the forth diagrams in $e\bar{e}$ sector,
together with the perturbative theory contribution to the order $O(e^2)$
of $ee\ga$ vertices (in the same sector), to calculate 
the mass of the positronium bound state. The rules to write the expression
of perturbative expansions from diagrams are given in \cite{ZhHa}.

Note, that the similarity function $f_{p_i p_j \la}$ in each term
restricts the energy differences between initial and final states
to be below the UV cutoff $\la$, and thus restricts the renormalized Hamiltonian 
to act in the low energy sector. Therefore, for the UV cutoff $\la<<\La$, we
associate the renormalized Hamiltonian with the effective
Hamiltonian descibing the physics of low energy. 

Another aspect of the flow equation method is connected with renormalization
group. Namely, the function $f_{p_ip_f\la}$ in the $ee\ga$-vertex plays 
the role of UV (and partialy IR) regulator in the self energy integrals
(see before), so that the regularization prescription of divergent 
integrals originates from the method of flow equations itself.
Moreover, as we will further show, the energy correction (i.e. mass correction
and wave function renormalization constant), obtained from the flow equation
method, coincide up to the overall sign with the $1$-loop renormalization group
result. This is the remarkable result, indicating the equivalence of
flow equations and Wilson's renormalization.

At last we mention that in the diagrammatic rules given in \fig{feynrules}
we explicitly write the dependence of the electron (photon) mass
on the cutoff, namely
\be
m_{\la}^2=m_0^2-\de\Si_{\la}
\; , \lee{ri44}
where $\de\Si_{\la}$ is the self energy term to the order $O(e^2)$,
($m_0=0$ for a photon). Whereas we drop the subscript $\la$ for the
polarization vectors $\varepsilon$ and spinors $\chi$. To the next order
$O(e^3)$ one has $e_{\la}=e_0^2(1+O(e_{\la}^2))$.

In the next section we use these diagrammatic rules of LFQED 
renormalized Hamiltonian, \fig{feynrules}, to calculate perturbatively 
electron-positron interaction, namely the contribution from 
perturbative photon exchange and electron (photon) self energy correction, 
which latter enables to find the corresponding physical masses.

\section{Positronium's fine structure}

\subsection{LF Perturbative theory}

The scattering $|e\bar{e}>$ states are also needed 
in bound state calculations.  Using the propagator
techniques we include these states where required.
We exploit the perturbative theory in the coupling constant
$e$, using the Feynman rules of the renormalized theory \fig{feynrules}.

The first order renormalized $ee\ga$-vertex \mbox{$f_{p_ip_f}H_{can}^{ee\ga}$}
contributes to the second order
to the $|e\bar{e}>$ interaction term and to the 
electron (photon) mass renormalization.
Physically, it is the perturbative photon exchange 
(photon emission and absorbtion in the case of electron mass renormalization),
with the energy widths of the photon restricted by 
the function $f_{p_ip_f,\la}$.

\subsubsection{The electron-positron interaction}

According to the light-front Feynman rules
the perturbative photon exchange gives rise to the following
second order $|e\bar{e}>$ interaction in the exchange channel 
\be
V(l)=g_1(l)g_2(l)M_{2ii} \cdot
\left\{ \frac{\th(q^+)}{q^+}\frac{1}{p_i^--p_k^-}
     + \frac{\th(-q^+)}{(-q^+)}\frac{1}{p_i^--p_k^-} \right\}
\; , \lee{epi1}
where $g_i$ stands schematically for the coupling constants in both vertices,
namely $g_{p_1p_2\la}=ef_{p_1p_2\la}$ and $M_{2ii}$ defines the spin structure of 
the interaction, coming from the corresponding structure of the $ee\ga$-vertex;
and the two terms in the curly brackets represent two different $x^+$ (time) orderings
of the photon exchange with the momentum $q$, giving rise to the two different
intermediate states with momenta $p_k$, $p_i$ corresponds to the initial state.
Explicitly one has

In the {\bf exchange channel} 
\bea
\hspace{0em} V_{PT}^{(ex)}(l) &=& -e^2(f_{-p_4,-p_2}(l)
\ch_{\bar{s}_2}^+\Ga^i(-p_4,-p_2,-q)\ch_{\bar{s}_4})
(f_{p_1,p_3}(l)
\ch_{s_3}^+ \Ga^i(p_1,p_3,q)\ch_{s_1}) \\
&&\hspace{2em} \times \left[ \frac{\th(p_1^+-p_3^+)}{(p_1^+-p_3^+)}\frac{1}{p_i^--p_3^--p_2^--q^-}
+\frac{\th(p_3^+-p_1^+)}{(p_3^+-p_1^+)}\frac{1}{p_i^--p_1^--p_4^-+q^-} \right] \,
\bar{\de}_{q,p_1-p_3} \nn
\leea{epi2}
with the initial state momentum $p_i=P=(P^+,P^{\bot})$ and momentum transfer
$q=p_1-p_3$, and
\be
P^-=\frac{P^{\bot 2}+M_N^2}{P^+}
\; , \lee{epi3}
where $M_N$ is the mass of positronium bound state.
In the light-front frame
\bea
-(P^--p_3^--p_2^--(p_1-p_3)^-)\th(p_1^+-p_3^+)
&=&(P^--p_1^--p_4^-+(p_1-p_3)^-)\th(p_3^+-p_1^+)\nonumber\\
&=&\frac{\tilde{\De}_3}{P^+(x-x')}
\leea{epi4}
holds, giving for the rescaled potential $V\rightarrow P^{+ 2}V$ rise to
\be
V_{PT,\la}^{(ex)} = -e^2N_1 \frac{1}{\tilde{\De}_3}
\exp\left( -\frac{(\De_1^2+\De_2^2)}{\la^4} \right)
\; , \lee{epi5}
where $N_1$ is defined in \eq{gi16} and
\bea
\tilde{\De}_3=(k_{\bot}-k'_{\bot})^2 + \frac{1}{2}(x-x')A+|x-x'|
\left( \frac{1}{2}(M_0^2+M_0^{'2})-M_N^2 \right)
\nonumber\\
\\
A=(k_{\bot}^2+m^2) \left( \frac{1}{1-x}-\frac{1}{x} \right)
+(k_{\bot}^{'2}+m^2) \left(\frac{1}{x'}-\frac{1}{1-x'} \right)
\nonumber
\; , \leea{epi6}
Here the cutoff $\la$ is defined in units of $P^+$.

Because of the absence of Z-graphen in light-front formalism (corresponding to
negative $p^+$), only one term contribute to the {\bf annihilation channel},
namely  

\bea
V_{PT}^{(an)}(l)&=&e^2(f_{p_1,-p_2}(l)
\ch_{\bar{s}_2}^+\Ga^i(p_1,-p_2,q)\ch_{s_1})
(f_{-p_4,p_3}(l)
\ch_{s_3}^+\Ga^i(-p_4,p_3,-q)\ch_{\bar{s}_4})\nonumber\\
&& \times \left[ \frac{1}{(p_1^++p_2^+)}\frac{1}{p_i^--q^-} \right]
\bar{\de}_{q,p_1+p_2}
\; , \leea{epi7}
where $p_i^-=P^-$ and the momentum transfer is $q=p_1+p_2$. This gives rise
for the rescaled potential $V\rightarrow P^{+2}V$ in the light-front frame,
to the expression
\be
V_{PT,\la}^{(an)} = e^2N_2\frac{1}{M_N^2}
\exp\left\{ -\frac{(M_0^4+M_0^{'4})}{\la^4} \right\}
\; , \lee{epi8}
where $N_2$ and the variables $\De_1, \De_2$ and $M_0^2, {M'}_0^2$
are defined in \eq{gi16}.

\subsubsection{Mass renormalization}

Following light-cone rules the perturbative energy correction
of the electron with momentum $p$, coming from the emission and 
absorption of a photon with momentum $k$, is
\bea
&& \de \tilde{p}_{1\la}^-=\int\frac{d^2k^{\bot}dk^+}{2(2\pi)^3}\frac{\th(k^+)}{k^+}
\th(p^+-k^+)g_{p-k,p,\la}\Ga^i_{\la}(p-k,p,-k)
g_{p,p-k,\la}\Ga^i_{\la}(p,p-k,k)\nn\\
&&\hspace{5em} \times\frac{1}{p^--k^--(p-k)^-}
\; , \leea{mr1}
where $g_{ee\ga}$-coupling constant restricts the energy of the
photon.
Making use of the explicit form for the coupling, one has
\bea
\de \tilde{p}_{1\la}^- &=& e^2\int
 \frac{d^2k^{\bot}dk^+}{2(2\pi)^3}\frac{\th(k^+)}{k^+}
 \th(p^+-k^+) \\
&&\hspace{3em} \times \Ga^i_{\la}(p-k,p,-k) \,
 \Ga^i_{\la}(p,p-k,k)\frac{1}{p^--k^--(p-k)^-}
 \times (R) \nn
\; , \leea{mr1a}
where $R=f_{pk\la}^2$ plays the role of regulator. 
This expression coincide up to the overall sign with
the energy correction obtained in the previous section from the flow equations method.

Two instantaneous diagramms, arising from the normal-ordering Hamiltonian,
must be added to the first term with the same regulator $R$.
Then the full perturbative energy correction
$\de\tilde{p}_{\la}^-=\de\tilde{p}_{1\la}^-+\de\tilde{p}_{2\la}^-+
\de\tilde{p}_{3\la}^-$ is
\be
\de\tilde{p}_{\la}^-=-\de p_{\la}^-
\ee
where $\de p_{\la}^-$ is defined in \eq{ri6}. This means for the perturbative mass
correction
\be
\de m_{\la}^{PT2}=\de\Sigma_{\la}
\lee{mr1b}
and the self-energy term $\de \Sigma_{\la}$ is given in \eq{ri24}.

We combine the renormalized to the second order mass, \eq{ri5},
and the perturbative correction, \eq{mr1b}, to obtain
the total physical mass to the order $O(e^2)$ 
\be
m_e^2=m_{\la}^2+\de m^2=(m^2+\de\Sigma_{\la})-\de\Sigma_{\la}\\
=m^2+O(e^4)
\; . \lee{mr3}
This means, that to the second order $O(e^2)$ the physical electron mass is,
up to a finite part, equal to the bare electron mass, that stands 
in the free (canonical) Hamiltonian.

Along the same line one can do for the photon mass.

\subsection{Bound state perturbative theory}

In this subsection we define bound state perturbative theory (BSPT).

First introduce instead of the front parametrization, used before
for the single-particle momenta, \fig{reneebarint}, the instant form
\bea
& p_{1\mu}=(xP^+,xP^{\bot}+k^{\bot},p_1^-) \quad
\stackrel{J(p)}{\longrightarrow} \quad
p_{1\mu} = (k_z, k^{\bot},p_1^0)=(\vec{p}_1,E_1) & \nn \\
& p_{2\mu}=((1-x)P^+,(1-x)P^{\bot}-k^{\bot},p_2^-) \quad
\longrightarrow \quad
p_{2\mu} = (-k_z, -k^{\bot},p_2^0)=(\vec{p}_2,E_2) & \nn \\
& E_i=\sqrt{\vec{p}^2+m^2}, \quad i=1,2 & \nn\\
& x =\frac{E_1+k_z}{E_1+E_2}= \frac{1}{2} \, \left( 
1 + \frac{k_z}{\sqrt{\vec{p}^2 + m^2}} 
\right) & \nn \\
\leea{bspt1}
and for the momenta $p_3,p_4$ the same, but with prime over
\mbox{$x, k_z, k^{\bot}$}; here $x$ is the light-front fraction of the
electron momentum, and $J(p)$ is the Jacobian of the transformation:
\be
J(p)=\frac{dx}{d k_z}=\frac{k_{\bot}^2+m^2}{2(\vec{p}^2+m^2)^{3/2}}
\; . \lee{bspt2}
The new coordinate system corresponds to the center of mass (c.m.) 
frame, i.e. $\vec{P}=0$.
Note, that in both sytems electron and positron are moving along paths
on the light-cone (against each other in the instant case
\mbox{$P_3=k_{z1}+k_{z2}=0$}), while the connection between the two
frames is obtained by boost.

The light-front bound state equation (see later) is boost and frame invariant;
therefore it can be solved in the c.m. frame, producing the same spectrum as 
in the front form. We shall see that the electron-positron potential,
arising from the renormalized to the second order Hamiltonian, is drastically
simplified in the c.m. instant frame. Also, this system is usefull in practical
aspect: in making obvious the rotational symmetry, restored in the nonrelativistic
limit and manifest in the spectrum. 

We stress an important consequence of this change of coordinates.
Since the total light-front momentum \mbox{$P^+ = P_0 + P_3$} is conserved
(i.e. both `in` \mbox{$|x,k^{\bot}>$} and `out` \mbox{$|x',k'^{\bot}>$}
Fock states have the same $P^+$) and $P_3=0$ in instant form, we have
{\it conservation of the total energy} $P_0 = P'_0$, or, in other words,
\mbox{$E_1 + E_2 = E_3 + E_4$}. For the $3-d$ momentum this means
\be
\vec{p}^{\,2} = \vec{p'}^2
\; . \lee{bspt3}
This condition we will further exploite in what follows.  

Define now {\bf BSPT}. We choose the leading order 
electron-positron potential in such a form to simplify positronium 
bound state calculations. This means, that this potential contributes 
the leading order term to the positronium mass, and perturbative 
theory with respect to the difference between the total
second order $|e\bar{e}>$ interaction, calculated before with 
the renormalized Hamiltonian $H_r$ to $O(e^2)$, and the leading order 
potential converges. This scheme we call BSPT. Surely, our choice 
is motivated by the form of the renormalized to the second order 
interaction to insure convergence of BSPT.  

We define the second order renormalized electron-positron potential
\mbox{$\left. \left< \! e(3)\bar{e}(4) \right|
\hat{V}_{coul} \left| e(1)\bar{e}(2) \! \right> \right.$}
to the leading order of BSPT in the form of pure perturbative one photon
exchange, explicitly as the Coulomb interaction 
\be
V_{coul} = -\frac{16 e^2 m^2}{(k_{\bot} - k'_{\bot})^2 + \\
(k_z - k'_z)^2} = -\frac{16 e^2 m^2}{(\vec{p}-\vec{p'})^2}
\; . \lee{bspt4}
This means that the corresponding leading order Hamilton operator 
in the \mbox{$|e\bar{e}>$} sector is
\be
H^{(0)} = h + \hat{V}_{coul}
\; , \lee{bspt5}
where $h$ is the free part, defined in \eq{ch13}. The wave functions are given
as the solution of Schr\"odinger equation
\be
H^{(0)} |\psi_N(P)> = E_N |\psi_N(P)>
\; , \lee{bspt6}
where $P$ is the positronium momentum, and the eigenvalues and 
eigenfunctions for the positronium bound state are defined in standard way
of light front frame
\bea
& E_N = \frac{P_{\bot}^2 + M_N^2}{P^+} & \nn \\
& |\psi_N(P)> = \sum_{s_1 s_2} \, \int_{p_1 p_2} \, \sqrt{p_1^+ p_2^+} \, 
2(2\pi)^3 \, \de^{(3)}(P - p_1 p_2) \, \tilde{\Phi}_N (x k_{\bot} s_1 s_2) \,
b^+_{s_1}(p_1) \, d^+_{s_2}(p_2)|0> & \nn \\
& \sum_{s_1s_2} \, \frac{\int d^2k_{\bot} \, \int_0^1 dx}{2(2\pi)^3} \,
\tilde{\Phi}_N^*(x k_{\bot} s_1 s_2) \,
\tilde{\Phi'}_N(x k_{\bot} s_1 s_2) = \de_{NN'} &
\; ; \leea{bspt7}
$M_N$ stands for the leading order mass of positronium.
Combining the definitions for the wave function and the energy 
with the Schr\"odinger equation, we obtain
\be
\biggl[ M_N^2-\frac{k_{\bot}^{'2}+m^2}{x'(1-x')} \biggr]
\tilde{\Phi}_N (x'k'_{\bot}s_3s_4) = \sum_{s_1 s_2}
\frac{\int d^2k_{\bot} \int_0^1 dx}{2(2\pi)^3} \,V_{coul} \, 
\tilde{\Phi}_N(xk_{\bot}s_1s_2)
\; , \lee{bspt8}
or, after change of coordinates according to \eq{bspt1},
\be
\left( M_N^2 - 4 (\vec{p'}^2 + m^2) \right) \Phi_N(\vec{p'} s_3 s_4) =
\sum_{s_1 s_2} \frac{\int d^3p \sqrt{J(p) J(p')}}{2(2\pi)^3} \,
V_{coul}(\vec{p},\vec{p'}) \, \Phi_N(\vec{p} s_1 s_2)
\; , \lee{bspt9}
where the wave function was redefined to have the norm
\be
\sum_{s_1s_2} \, \int d^3p \, \Phi_N^*(\vec{p} s_1 s_2) \,
\Phi_N'(\vec{p} s_1 s_2) = \de_{NN'} 
\; . \lee{bspt10}
We aim to obtain the nonrelativistic Schr\"odinger equation for positronium.
Note, that in the nonrelativistic limit \mbox{$\frac{\vec{p}^2}{m^2} <\!\!< 1$}
we have 
\bea
& \sqrt{J(p) J(p')} \; \approx \;
\frac{1}{2m} \left( 1 - \frac{\vec{p}^2 + (k_z^2 + {k'_z}^2)} {2m^2} \right) & \nn \\
& M_N \; = \; (2m + B_N)^2 \; \approx \; 4m^2 + 4m B_N^{(0)} &
\; , \leea{bspt11}
where we have introduced the leading order binding energy $B_N^{(0)}$.
Then to the leading order the bound state equation for positronium is
\bea
& \left( \frac{\vec{p'}^2}{m} - B_N \right) \Phi_N(\vec{p'} s_3 s_4)
\; = \;- \sum_{s_1 s_2} & \nn \\
& \int d^3p \left( \frac{1}{2m} \frac{1}{2(2\pi)^3} \frac{1}{4m} \, V_{coul} \right) \,
\Phi_N(\vec{p}s_1s_2) &
\; . \leea{bspt12}

Making use of the explicit form for the Coulomb potential, \eq{bspt4}, we obtain
the equation that determines the leading order bound state wave function:
\be
\left( \frac{\vec{p'}^2}{m} - B_N \right) \, \Phi_{\mu}(\vec{p'}) \; = \;
\frac{\al}{2\pi^2} \, \int \, \frac{d^3p}{(\vec{p}-\vec{p'})^2} \, \Phi_{\mu}(\vec{p})
\lee{bspt13}
with
\be
\Phi_N \; = \; \Phi_{\mu, s_e, s_{\bar{e}}}(\vec{p'} s_3 s_4) \; = \;
\Phi_{\mu}(\vec{p'}) \, \de_{s_e s_3} \, \de_{s_{\bar{e}} s_4}
\; . \lee{bspt14}

This is the standard nonrelativistic Schr\"odinger equation 
for positronium. Its solution is characterized by
\mbox{$\mu \! = \!(n,l,m)$}, the usual principal and angular momentum quantum numbers.
The wave functions are given through the hyperspherical harmonics
\bea
&& Y_{\mu}(\Om) \; = \;
\frac{(e_n^2 + \vec{p}^2)^2}{4 \, e_n^{5\!/\!2}} \, \Phi_{\mu} \nn \\
&& Y_{\mu} \; = \; Y_{n,l,m} \; = \; f_{n,l}(\om) \, Y_{l,m}(\theta,\phi) \nn \\
&& B_N \; = \; -\frac{m \alpha^2}{4n^2},\qquad e_n \; = \; \frac{m\alpha}{2n} 
\leea{bspt15}
and for the binding energy one has the standard 
nonrelativistic expression for positronium bound state to $O(e^2)$.
For sake of completeness we write the coordinates used in the solution
\bea
&& (e_n^2 = -m B_N, \vec{p}) \; \longrightarrow \; (u_0, \vec{u}) \nn \\
&& u_0 \; = \; \cos\om \; = \; \frac{e_n^2 - \vec{p}^2}{e_n^2 + \vec{p}^2} \nn \\
&& \vec{u} \; = \; \frac{\vec{p}}{|\vec{p}|}
\sin\omega \; = \; \frac{2e_n \vec{p}}{e_n^2 + \vec{p}^2} 
\; , \leea{bspt16}
but, for details, refer to \cite{JoPeGl}.

The electron-positron interaction arising from the renormalized
to the $O(e^2)$ Hamiltonian is given as a sum of two contributions
from exchange and annihilation channels \mbox{$V=V_{exch}+V_{ann}$}
(see explicitly later). We introduce the potential, arising in the
nonrelativistic Schr\"odinger equation, \eq{bspt13},
\be
\widetilde{V}(\vec{p}^{'}s_3s_4;\vec{p}s_1s_2) =
\lim_{\frac{\vec{p}^2}{m^2}<<1} \frac{\sqrt{J(p)J(p')}}{2(2\pi)^3}\frac{1}{4m}
(V_{exch}+V_{ann})
\; . \lee{bspt17}
Then we define BSPT with respect to the difference
\be
\de V=\widetilde{V}(\vec{p'}s_3s_4;\vec{p}s_1s_2)
-(-\frac{\alpha}{2\pi^2})\frac{1}{(\vec{p}-\vec{p}^{'})^2}
\de_{s_1s_3}\de_{s_2s_4}
\; , \lee{bspt18}
where the leading order contribution is defined in \eq{bspt15}.
Note, that, to define the Coulomb potential, we have taken only the first
term of the Jacobian's $J(p)$ nonrelativistic expansion, i.e.
the $e\bar{e}$ interaction to the leading order of BSPT.

In what follows we use the matrix elements of $\de V$, defined as 
\bea
&& <\Phi_{nlm}|\de V|\Phi_{nlm}>=\int d^3pd^3p'
\Phi_{nlm}^*(\vec{p})\de V\Phi_{nlm}(\vec{p'})
\; , \leea{bspt19}
where $\Phi_{nlm}$ are the Coulomb wave functions given above.

\subsection{The re\-normalized elec\-tron-po\-si\-tron interaction in light-front
 and instant form frames}

We summarize all together the second order $| e\bar{e}\! >$ interaction
in the exchange and annihilation channels, i.e., written in the 
{\bf front form frame}: 
\bea
V_{exch} &=& V_{\la}^{exch} + V^{PT}
 \;=\; V_{\la}^{gen} + V_{\la}^{inst} + V_{\la}^{PT} \nn \\ 
V_{ann} &=& V_{\la}^{ann} + V^{PT}
 \;=\; V_{\la}^{\prime \: gen} + V_{\la}^{\prime \: inst} + 
V_{\la}^{\prime \: PT}
\; , \leea{repi1}
where the generated, instantaneous and perturbative theory interactions
(rescaled, i.e. $V\rightarrow P^{+ 2}V$) are given correspondingly 
in the {\bf exchange channel}
\bea
V_{\la}^{gen} &=& -e^2 N_{1,\la} \cdot \frac{1}{2}
 \left(\frac{1}{\widetilde{\De}_1} + \frac{1}{\widetilde{\De}_2} \right)
 \left(1 - e^{-2 \frac{\De_1}{\la^2} \cdot \frac{\De_2}{\la^2}} \right)
 e^{-\left(  \frac{M^2_0 - M'^2_0}{\la^2} \right)^2 } \nn \\
V_{\la}^{inst} &=& -\frac{4 e^2}{(x - x')^2} \: \de_{s_1 s_3} \: 
\de_{s_2 s_4} \cdot
 e^{-\left(  \frac{M^2_0 - M'^2_0}{\la^2} \right)^2 } \nn \\
V_{\la}^{PT} &=& -e^2 N_{1,\la} \cdot \frac{1}{\widetilde{\De}_3}
 e^{- \left( \left( \frac{\De_1}{\la^2} \right)^2 +  \left( 
\frac{\De_2}{\la^2} \right)^2 \right)}
\leea{repi2}
in the {\bf annihilation channel}
\bea
V_{\la}^{gen} &=& e^2 N_{2,\la} \cdot \frac{1}{2}
 \left(\frac{1}{M_0^2} + \frac{1}{M'^2_0} \right)
 \left(1 - e^{-2 \frac{M^2_0}{\la^2} \cdot \frac{M'^2_0}{\la^2} } \right)
 e^{-\left(  \frac{M^2_0 - M'^2_0}{\la^2} \right)^2 } \nn \\
V_{\la}^{inst} &=& 4 e^2
 \: \de_{s_1 \bar{s}_3} \: \de_{s_2 \bar{s}_4} \cdot
 e^{-\left(  \frac{M^2_0 - M'^2_0}{\la^2} \right)^2 } \nn \\
V_{\la}^{PT} &=& e^2 N_{2,\la} \cdot \frac{1}{M_{\!N}^2}
 e^{- \left( \left( \frac{M^2_0}{\la^2} \right)^2 +  \left( 
\frac{M'^2_0}{\la^2} \right)^2 \right)}
\; , \leea{repi3}
where in the light-front frame, see \eq{b14} in Appendix B,
\bea
N_{1,\la}&=&\delta_{s_1s_3}\delta_{s_2s_4}
 T_1^{\bot}\cdot T_2^{\bot}
 -\delta_{s_1\bar{s}_2}\delta_{s_1\bar{s}_3}\delta_{s_2\bar{s}_4}
 2m^2\frac{(x-x')^2}{xx'(1-x)(1-x')}\nonumber\\
&&+im\sqrt{2}(x'-x) \left[ \delta_{s_1\bar{s}_3}\delta_{s_2s_4}
 \frac{s_1}{xx'}T_1^{\bot}\cdot \varepsilon_{s_1}^{\bot}
 +\delta_{s_1s_3}\delta_{s_2\bar{s}_4}
 \frac{s_2}{(1-x)(1-x')}T_2^{\bot}\cdot \varepsilon_{s_2}^{\bot} \right]
\nn\\
N_{2,\la}&=&\delta_{s_1\bar{s}_2}\delta_{s_3\bar{s}_4}
 T_3^{\bot}\cdot T_4^{\bot}
 +\delta_{s_1s_2}\delta_{s_3s_4}\delta_{s_1s_3}
 2m^2\frac{1}{xx'(1-x)(1-x')}\nonumber\\
&&+im\sqrt{2} \left[ \delta_{s_3\bar{s}_4}\delta_{s_1s_2}
 \frac{s_1}{x(1-x)}T_3^{\bot}\cdot \varepsilon_{s_1}^{\bot}
 -\delta_{s_3s_4}\delta_{s_1\bar{s}_2}
 \frac{s_3}{x'(1-x')}T_4^{\bot}\cdot \varepsilon_{s_4}^{\bot *} \right] \nn \\
&& \varepsilon_s^i = -\frac{1}{\sqrt{2}}(s, i)
\leea{repi4}
and
\bea
T_1^i&=&- \left[ 2\frac{(k_{\bot}-k'_{\bot})^i}{(x-x')}+\frac{k_{\bot}^i(s_2)}{(1-x)}+
 \frac{k_{\bot}^{'i}(\bar{s}_2)}{(1-x')} \right] \; ; \qquad
T_2^i=2\frac{(k_{\bot}-k'_{\bot})^i}{(x-x')}-\frac{k_{\bot}^i(s_1)}{x}-
 \frac{k_{\bot}^{'i}(\bar{s}_1)}{x'} \nonumber \\
T_3^i&=&-\frac{k_{\bot}^{'i}(\bar{s}_3)}{x'}
 +\frac{k_{\bot}^{'i}(s_3)}{(1-x')} \; ; \qquad
T_4^i=\frac{k_{\bot}^i(\bar{s}_1)}{(1-x)}
 -\frac{k_{\bot}^i(s_1)}{x}\nonumber\\
&& k_{\bot}^i(s) = k_{\bot}^i+is\varepsilon_{ij}k_{\bot}^j \; ; \qquad
 \varepsilon_{ij}=\varepsilon_{ij3} \; ; \qquad
 \bar{s} = -s  \nn
\leea{repi5}
with the definitions
\bea
\tilde{\De}_1 &=& \frac{(xk'_{\bot}-x'k_{\bot})^2+m^2(x-x')^2}{xx'}\; ; \qquad
 \tilde{\De}_2=\De_1|_{x\rightarrow(1-x),x'\rightarrow(1-x')} \nn \\
&&\De_1=\frac{\tilde{\De}_1}{x'-x} \; ; \qquad
 \De_2=\frac{\tilde{\De}_2}{x'-x} \nn \\
\tilde{\De}_3 &=& (k_{\bot}-k'_{\bot})^2+\frac{1}{2}(x-x')A+|x-x'|
 \left( \frac{1}{2}(M_0^2+M_0^{'2})-M_N^2 \right) \nn \\
&&M_0^2=\frac{k_{\bot}^2+m^2}{x(1-x)} \; ; \qquad
 M_0^{'2}=\frac{k_{\bot}^{'2}+m^2}{x'(1-x')} \nn \\
&&A=(k_{\bot}^2+m^2) \left( \frac{1}{1-x}-\frac{1}{x} \right)
 +(k_{\bot}^{'2}+m^2) \left( \frac{1}{x'}-\frac{1}{1-x'} \right) \nn \\
&&P^-=\frac{(P^{\bot})^2+M_N^2}{P^+} \; ; \qquad P=(P^+,P^{\bot}) \; ; \qquad
 M_N = 2m + B_N
\; . \leea{repi6}
Note, that the rescaled potential, \eq{repi1}, does not depend on the
total momentum $P^+$, i.e. is invariant under light-front boosts.

The generated interaction plays an important role,
namely it insures the absence of collinear divergencies
in the renormalized interaction, which are associated with 
the limit $x\longrightarrow x'$. This is true for any
cutoff $\la$. In fact, in the limit when $x$ tends to $x'$ one has
\bea
&& \tilde{\De}_1\sim\tilde{\De}_2\longrightarrow
(k_{\bot}-k_{\bot}^{'})^2 \nn\\
&& \De_1\sim\De_2\longrightarrow
\frac{(k_{\bot}-k_{\bot}^{'})^2}{x'-x}
\; . \leea{repi7}
This gives rise to
\bea
&& V^{exch}_{\la}(x\rightarrow x')=\left[ -\frac{e^2N_{1,\la}}
{(k_{\bot}-k_{\bot}^{'})^2}c_{ex}^{gen}-\frac{4e^2}{(x-x')^2}
c_{ex}^{inst} \de_{s_1s_3}\de_{s_2s_4} \right]
e^{-\left(  \frac{M^2_0 - M'^2_0}{\la^2} \right)^2 }
\; , \leea{repi8}
where, in this limit, the perturbative part vanishes.
Making use of \eq{repi4} we obtain for the divergent part ${\cal DP}[\cdots]$
\bea
&& {\cal DP}[N_{1,\la}(x\rightarrow x')]=T_1^{\bot}T_2^{\bot}
\de_{s_1s_3}\de_{s_2s_4}\nn\\
&& {\cal DP}[T_1^i(x\rightarrow x')]=-2\frac{(k_{\bot}-k_{\bot}^{'})^i}
{(x-x')} \; ;\quad
{\cal DP}[T_2^i(x\rightarrow x')]=2\frac{(k_{\bot}-k_{\bot}^{'})^i}
{(x-x')}
\; . \leea{repi9}
This results in the exact cancellation of the collinear divergencies
present initially in the instantaneous interaction by the generated term:
\be
{\cal DP}[V^{exch}_{\la}(x\rightarrow x')]=0
\; . \lee{repi10}
Note, that {\it no approximation} for the $e\bar{e}$ interaction 
was made untill now.

We rewrite both exchange and annihilation channel contributions
in the {\bf instant frame}
\bea
V &=& V_{exch}+V_{ann} \nn \\
&=&-e^2N_{1,\la} \left[ \frac{1}{2}(\frac{1}{\tilde{\De}_1}
 +\frac{1}{\tilde{\De}_2})
 (1-e^{-2 \frac{\De_1}{\la^2} \cdot \frac{\De_2}{\la^2}})c_{ex}^{gen}
 +\frac{1}{\tilde{\De}_3}
 e^{- \left( \left( \frac{\De_1}{\la^2} \right)^2 +  \left( 
 \frac{\De_2}{\la^2} \right)^2 \right)}c_{ex}^{PT} 
 \right] \nn \\
&&+\left( -\frac{4 e^2}{(x - x')^2} \: \de_{s_1 s_3} \: 
 \de_{s_2 s_4} \right)c_{ex}^{inst}\nn\\
&&+e^2N_{2,\la} \left[ \frac{1}{M_0^2}
 \left(1 - e^{-2 (\frac{M_0^2}{\la^2})^2} \right)c_{an}^{gen}
 +\frac{1}{M_N^2}
 e^{- 2\left( \frac{M_0^2}{\la^2} \right)^2}c_{an}^{PT}
 \right] \nn \\
&&+\left( 4 e^2
 \: \de_{s_1 \bar{s}_3} \: \de_{s_2 \bar{s}_4} \right)c_{an}^{inst}
\; , \leea{repi11}
where quantities in the instant form are defined as follows
\bea
x&=&\frac{1}{2}\left(1+\frac{k_z}{\sqrt{\vec{p}^2+m^2}}\right) \; ;\qquad
x'=\frac{1}{2}\left(1+\frac{k_z^{'}}{\sqrt{\vec{p}^2+m^2}}\right) \nn \\
M_0^2&=&M_0^{'2} = 4(\vec{p}^2+m^2)
\; ; \leea{repi12}
for the other quantities, defined in \eqs{repi4}{repi5} the substitution 
$x(k_z), x'(k_z^{'})$ is to be done. The symbols $c_{ex}^{gen}$
and so on were introduced to indicate the origin of the different terms 
(here generated interaction coming from the exchange channel), all $c=1$. 
We stress once more, that at the moment no approximation 
was done, therefore the expression of \eq{repi11} can be considered as the
{\it exact electron-positron interaction, arising from the renormalized
to the second order Hamiltonian}, that produces to the leading order
the positronium mass $M_N$ and the binding energy $B_N$. 
In fact, it will be shown, that the second order potential
gives rise also to the correct mass splitting, that corresponds
to the next to leading order contribution.

The expression of \eq{repi11} has a transparant form. First notice, that
$\De_1$ and $\De_2$ descibe the energy differences in two corresponding
$ee\ga$-vertices appearing in the $e\bar{e}$ interaction. Then
the generated interaction ($c_{ex}^{gen}$) contributes
mainly hard photon exchanges
\mbox{$\frac{\De_1}{\la^2} \sim \frac{\De_2}{\la^2}>>1$},
while the term arising from perturbative
theory $c_{ex}^{PT}$ gives rise to soft photon exchanges. Though
the renormalized interaction generally descibes low energy 
physics, namely the renormalized Hamiltonian acts in the space
where the energy differences are restricted by $\la$, the information
on the high energy sector is accumulated in the generated interaction,
making possible to interpolate between two sectors.
This means, that the sum of both terms in \eq{repi11} recovers the
{\it whole range of photon energies}.
The same is true for the annihilation channel.

Two limits are of interest. The trivial limit 
$\la\longrightarrow\La$ reproduces the bare interaction,
used as initial condition for the renormalized interaction
as $\La$ tends to infinity. In the opposite limit
\be
\la <\!\!< m
\lee{repi13}
nonrelativistic physics is affected (see later).
The limit $\la\longrightarrow 0$ can be performed {\it explicitly},
corresponding to the complete elimination of the $ee\ga$-vertex present
in the initial Hamiltonian. When $\la$ tends to zero the $e\bar{e}$
interaction is governed by generated and instantaneous terms.
The fact that in this limit the potential is well defined
is the direct consequence of the instant form frame.

The expression of \eq{repi11} for the renormalized
to the second order interaction, written in the instant form frame,
can be used as a suitable form to obtain the positronium spectrum numerically.
We further proceed in another direction, namely we make the nonrelativistic
approximation (NR) for the electron (positron) momentum
\be
\frac{|\vec{p}|}{m}=O(\alpha)<1
\; , \lee{repi14}
that enables to perform calculations for positronium mass 
splitting analitically.

\subsection{The nonrelativistic approximation}

In the nonrelativistic approximation  we obtain for the $e\bar{e}$
interaction, \eq{repi11},
\bea
\hspace{-6em} \tilde{V}_{\la} &=& \frac{1}{2(2\pi)^3} \frac{1}{4m} \frac{1}{2m} \,
 \biggl(1 - \frac{\vec{p}^2}{2m^2}\biggr) \, V_{\la} \nn \\
\hspace{-6em} V_{\la} &=& V_{\la}^{exch} + V_{\la}^{ann} \nn \\
&=& - \frac{e^2N_1}{(\vec{p} - \vec{p'})^2}
      \biggl(1 - {\rm e}^{-2 \left(\frac{\De}{\la^2} \right)^2} \biggr) \, c_{ex}^{gen} \,
     -\frac{e^2N_1}{(\vec{p}-\vec{p'})^2 + |x - x'|(M_0^2 - M_N^2)} \,
      {\rm e}^{-2 \left(\frac{\De}{\la^2}\right)^2} \, c_{ex}^{PT} \, \nn \\
 && - \: \frac{4 e^2}{(x - x')^2} \: \de_{s_1 s_3} \: 
      \de_{s_2 s_4} \, c_{ex}^{inst}\nn\\
 && + \: \frac{e^2N_2}{4m^2}
      \biggl( 1 - {\rm e}^{-2 \left(\frac{4m^2}{\la^2} \right)^2} \biggr) \,c_{an}^{gen}
     +\frac{e^2N_2}{M_N^2} \,
      {\rm e}^{-2 \left(\frac{4m^2}{\la^2}\right)^2} \, c_{an}^{PT} \, \nn \\
 && + \: 4 e^2
      \: \de_{s_1 \bar{s}_3} \, \de_{s_2 \bar{s}_4} \, c_{an}^{inst}
\; , \leea{nra1}
where the energy denominators and exponential factors were
simplified using
\bea
&& x - x' = \frac{k_z-k_z^{'}}{2m} \left[ 1 + \frac{\vec{p}^2}{2m^2} \right]
 + O \left( m^2 \left(\frac{p}{m}\right)^5 \right) \nn \\
&& \tilde{\De}_1 = \tilde{\De}_2 = (\vec{p} -\vec{p'})^2
 + O \left(m^2 \left(\frac{p}{m}\right)^5 \right) \nn \\
&& \tilde{\De}_3 = (\vec{p} - \vec{p'})^2
 + | x - x' | (M_0^2 - M_N^2)
 + O \left(m^2 \left(\frac{p}{m}\right)^4 \right) \nn \\
&& \De_1 = \De_2 = \frac{2m(\vec{p'} - \vec{p})^2}{(k_z^{'}-k_z)}
 \left[ 1 + O \left( \left(\frac{p}{m}\right)^2 \right) \right] \; ;\qquad
 \De = \frac{2m (\vec{p'} - \vec{p})^2}{(k_z' - k_z)} \nn \\
&& M_0^2 = 4m^2 + O \left(m^2 \left(\frac{p}{m} \right)^2 \right) \nn \\
&& M_N^2 = 4m^2 + 4m B_N + O \left(m^2 \left(\frac{B_N}{m}\right)^2 \right)
 = 4m^2 + 4m B_N^{(0)}
\; , \leea{nra2}
and the explicit expression of Jacobian for the coordinate change is
\be
\sqrt{J(p)J(p')} = \frac{1}{2m} \left[ 1 - \frac{\vec{p}^{\,2}}{2m^2}
 + O \left( \frac{k_z^2}{m^2}, \frac{k_z^{'2}}{m^2} \right) \right]
\; , \lee{nra3}
having introduced the leading order binding energy $B_N^{(0)}$.
Making use of its nonrelativistic approximation
$B_N^{(0)} \!/\! m <\!\!< 1$ we have for the interaction
\bea
&&\tilde{V}_{\la} = \frac{1}{2(2\pi)^3} \frac{1}{4m} \frac{1}{2m} \,
 \biggl( 1 - \frac{\vec{p}^{\,2}}{2m^2} \biggr) \nn \\
&&\hspace{3em} \times \biggl[ \biggl(
 - \frac{e^2 N_1}{(\vec{p} - \vec{p'})^2} \; (c^{gen}_{ex}, c^{PT}_{ex}) \; 
 - \frac{16 e^2 m^2}{(k_z - k'_z)^2} \,
 \left( 1 + \frac{\vec{p}^{\,2}}{m^2} \right) \; c^{inst}_{ex} \; 
 \de_{s_1 s_3}  \: \de_{s_2 s_4} \nn \\
&&\hspace{6.5em} + \; \frac{e^2 N_2}{4 m^2} \; (c^{gen}_{an}, c^{PT}_{an}) \;
 \hspace{8em} + \; 4e^2  \; c^{inst}_{an} \; \de_{s_1 \bar{s}_2} \: \de_{s_3 \bar{s}_4}
 \biggr) \nn \\
&&\hspace{3.5em} + \biggl(
 - \frac{e^2 N_1}{(\vec{p} - \vec{p'})^2} \frac{4m B_{\!N}^{(0)}}{|\De|} \;
 c^{PT}_{ex} \; {\rm e}^{- \left( \frac{\De}{\la^2} \right)^2 } \nn \\
&&\hspace{7.5em}
 - \; \frac{e^2 N_2}{4m^2} \frac{B_{\!N}^{(0)}}{m} \; c^{PT}_{an} \;
 {\rm e}^{- \left( \frac{4m^2}{\la^2} \right)^2 } \hspace{0.5em} \biggr)
 \biggr]
\; , \leea{nra4}
where $(c^{gen}, c^{PT})$ showes that both term, 
generated and perturbative interactions, contribute to corresponding
term (we remember that all $c = 1$).

The remarkable feature of the part of interaction standing
in the first bracket is that it does not depend on the UV cutoff $\la$.
The next term in the second bracket arises from the perturbative photon
exchange and has the typical 'energy shell' structure for the relativistic
effects, namely these terms are important when $\la>>m$. Further we
calculate the ground state positronium mass and therefore restrict
the cutoff to be in the nonrelativistic domain
\be
\la <\!\!< m
\; , \lee{nra5}
where the second term in \eq{nra4} vanishes and we are left with
the following form for the renormalized $e\bar{e}$ interaction
in the nonrelativistic approximation:
\bea
\hspace{-1em} \widetilde{V}(\vec{p},\vec{p'})
 &=& \frac{1}{2(2\pi)^3} \frac{1}{4m}\frac{1}{2m} \,
 \biggl( 1 - \frac{\vec{p}^2}{2m^2} \biggr) \\
&&\hspace{1em} \times \biggl[
 - \frac{e^2 N_1}{(\vec{p} - \vec{p'})^2} \; (c^{eff}_{ex}, c^{PT}_{ex}) \; 
 - \frac{16 e^2 m^2}{(k_z - k'_z)^2} \left( 1 + \frac{\vec{p}^2}{m^2} \right) \;
 c^{inst}_{ex} \; \de_{s_1 s_3} \: \de_{s_2 s_4} \nn \\
&&\hspace{3.8em} + \; \frac{e^2 N_2}{4m^2} \; 
 (c^{eff}_{an}, c^{PT}_{an}) \;
 \hspace{8.2em} + \; 4e^2 \; c^{inst}_{an} \; \de_{s_1 \bar{s}_2} \: \de_{s_3 \bar{s}_4}
 \biggr] \nn
\; . \leea{nra6}
We perform the nonrelativistic expansion of the factors $N_1$ and $N_2$
appearing in the interaction.

\noindent
The term $N_1$ contributes in $\widetilde{V}$ to the order

\noindent
\mbox{$\underline{O(1),O\left( \left(\frac{p}{m} \right)^2 \right) }$}:
\bea
-T_1^{\bot}T_2^{\bot} &=& 16m^2 \frac{q_{\bot}^2}{q_z^2}
 \left( 1 + \frac{\vec{p}^2}{m^2} \right) + 16 \frac{q_{\bot}^i}{q_z}
 \left( k_{\bot}^ik_z + k_{\bot}^{'i} k_z^{'} \right) \nn \\
&& -16i(s_1+s_2)[k_{\bot}^{'}k_{\bot}]-4(k_{\bot}+k_{\bot}^{'})^2
+4s_1s_2q_{\bot}^2\nn
\; , \leea{nra7}
 
\noindent
\mbox{$\underline{O\left(\frac{p}{m}\right),
 O\left( \left(\frac{p}{m} \right)^2 \right) }$} :
\bea
&&\hspace{-1em} im\sqrt{2}(x'-x)
 \left(\frac{s_1}{xx'} \: \varep^\perp_{s_1} \cdot T^\perp_1 \;
  \de_{\bar{s}_1 s_3} \: \de_{s_2 s_4} \;
  + \frac{s_2}{(1-x)(1-x')} \: \varep^\perp_{s_2} \cdot T^\perp_2 \;
  \de_{\bar{s}_4 s_2} \: \de_{s_1 s_3} \right) \nn \\
&&\hspace{0em} = 8 \, \de_{\bar{s}_1 s_3}  \: \de_{s_2 s_4} \,
 \left[m \, (iq_{\bot}^x - s_1q_{\bot}^y) \left( 1 - \frac{k_z+k_z^{'}}{m} \right) \,
  + q_z \, (i\tilde{p}_{\bot}^{x}-s_1\tilde{p}_{\bot}^{y})
  + \frac{1}{2} s_2 q_z(q_{\bot}^y - is_1q_{\bot}^x) \right] \nn \\
&&\hspace{1em} - \; \de_{\bar{s}_4 s_2} \: \de_{s_1 s_3}
 \left[m \, (iq_{\bot}^x - s_2q_{\bot}^y) \, \left( 1 + \frac{k_z + k_z'}{m} \right) \,
  - q_z \, (i\tilde{p}_{\bot}^{x}-s_2\tilde{p}_{\bot}^{y})
  - \frac{1}{2} s_1 q_z(q_{\bot}^y - is_2q_{\bot}^x) \right] \nn
\; , \leea{nra8}

\noindent
\mbox{\underline{$O\left( \left(\frac{p}{m} \right)^2 \right)$}} :
\bea
&& 2m^2\frac{(x-x')^2}{xx'(1-x)(1-x')} = 8 q_z^2 \nn
\; . \leea{nra9}

\noindent
Whereas the term $N_2$ contributes to $\widetilde{V}$ to the order

\noindent
\mbox{\underline{$O\left( \left(\frac{p}{m} \right)^2 \right)$}} :
\be
2m^2\frac{1}{xx'(1-x)(1-x')} = 32m^2 \nn
\; . \lee{nra10}

In these formulas we have used
\mbox{$[k_{\bot}^{'},k_{\bot}] = \varepsilon_{ij}k_{\bot}^{'i}k_{\bot}^{j}$},
$\varepsilon_{ij}=\varepsilon_{ij3}$ and
\mbox{$\varepsilon_{s}^{i} = -\frac{1}{\sqrt{2}}(s,i)$}; also
the following variables have been introduced
\bea
&& q_{\bot} = k_{\bot}^{'}-k_{\bot} \; ,\qquad (\bot=x,y)~,~
   q_z = k_z^{'}-k_z \nn \\
&& \tilde{p}_{\bot} = \frac{k_{\bot} + k'_{\bot}}{2}
\; . \leea{nra11}
We leave aside for the future work the analysis of the
expressions for $N_1$ and $N_2$, where also in this form some terms
can be identified as spin-orbit and spin-spin interactions
in the transverse plane and in longitudinal (z) direction.

Instead we follow \cite{JoPeGl}, where an analogous calculation
of singlet-triplet ground state mass splitting of positronium was performed 
in the similarity scheme. This means, that we can, except for the leading
order term $O(1)$, drop in $N_1$ the part diagonal in spin space. Also the
terms of the type 
$f=k_{\bot}^{x,y}k_z~,~k_{\bot}^{x,y}k_z^{'}~,~k_{\bot}^xk_{\bot}^y$
do not contribute to the ground state mass splitting, since
\bea
&& \int d^3pd^3p' \, \Phi_{100}^*(\vec{p}) \, 
\frac{f}{\vec{q}^2} \, \Phi_{100}(\vec{p'})
\; , \leea{nra12}
averaging over directions, gives zero.

We obtain for the $e\bar{e}$-potential to the leading order $O(1)$
of nonrelativistic expansion
\bea
\tilde{V}^{(0)}(\vec{p'}, \vec{p})
 &=& \frac{1}{2(2\pi)^3} \frac{1}{4m}\frac{1}{2m} \,
 \left( 1 - \frac{\vec{p}^2}{2m^2} \right) \nn \\
&& \times \left[ \frac{16e^2m^2}{\vec{q}^2} \frac{q_{\bot}^2}{q_z^2}
 \left( 1 + \frac{\vec{p}^2}{m^2} \right) \; (c_{ex}^{gen},c_{ex}^{PT}) \;
 - \frac{16e^2m^2}{q_z^2} \, \left( 1 + \frac{\vec{p}^2}{m^2} \right) \;
 c_{ex}^{inst} \; \right]
 \de_{s_1s_3} \: \de_{s_2s_4} \nn \\
&=& - \frac{\alpha}{2\pi^2} \frac{1}{\vec{q}^2} \left( 1 + \frac{\vec{p}^2}{2m^2} \right) \,
 \de_{s_1s_3} \: \de_{s_2s_4} \\
&&\hspace{-2em}\longrightarrow \quad V(r) \, \left( 1 + \frac{\vec{p}^2}{2m^2} \right) \nn   
\; . \leea{nra13}
Remembering \mbox{$\vec{q} = \vec{p'}-\vec{p}$}, Fourier transformation
to the coordinate space with respect to $\vec{q}$ has been performed
in the last expression. To the leading order of NR expansion
we have reproduced the Coulomb potential, defined before as 
the leading order of BSPT. Note, this is true for any UV cutoff
within the nonrelativistic range \mbox{$\la <\!\!< m$}.

We combine this expression with the kinetic term
from the Schr\"odinger equation, \eq{bspt13}, and write it in the form
\be
\frac{1}{m} \left( 1 + \frac{V(r)}{2m} \right) \vec{p}^{\,2} + V(r)
\; . \lee{nra14}
Here the potential $V(r)$ plays a different role in the two terms.
In the first term, corresponding to kinetic energy, it generates
an effective mass of the electron, which depends on the relative
position and manifests the non-locality of the interaction.
The second term is the usual potential energy, in our case, the Coulomb
interaction.

The energy of the Coulomb level with quantum numbers $(nlm)$ is standard
\be
M_0^2 = <\Phi_{nlm}|\tilde{V}^{(0)}|\Phi_{nlm}>
= \int d^3p d^3p'\Phi^{*}_{nlm}(\vec{p}) \, \tilde{V}^{(0)} \, \Phi_{nlm}(\vec{p'})
= -\frac{m\alpha^2}{2n^2}
\; , \lee{nra15}
where the Coulomb wave functions $\Phi_{nlm}$ were defined in \eq{bspt15}. 
We have used in \eq{nra15} the following representation
\bea
&& (\vec{p} - \vec{p'})^2 = \frac{(e_n^2 + \vec{p}^2) \,
(e_n^2 + \vec{p'}^2)}{4e_n^2} \, (u-u')^2 \nn \\
&& \frac{1}{(u-u')^2} = \sum_{\mu} \frac{2\pi^2}{n} \,
Y_{\mu}(\Omega_{p}) \, Y_{\mu}^{*}(\Omega_{p'}) \nn \\
&& d^3p = d\Omega_p \left( \frac{e_n^2 + \vec{p}^2}{2e_n} \right)^3
\leea{nra16}
and also orthogonality of the hyperspherical harmonics
\be
\int d\Omega \, Y_{\mu}^{*} \, Y_{\mu^{'}} = \de_{\mu\mu'}
\; . \lee{nra17}
More details can be found in \cite{JoPeGl}.

The next to leading order \mbox{$O\left(\frac{p}{m}\right)$}
\bea
\de V^{(1)} &=& \frac{1}{2(2\pi)^3} \frac{1}{4m} \frac{1}{2m}
 \left(-\frac{e^2}{\vec{q}^{\,2}}\right) \nn \\
&& \times \left(
  8m(iq_{\bot}^x-s_1q_{\bot}^y) \de_{s_1\bar{s}_3} \de_{s_2s_4}
 -8m(iq_{\bot}^x-s_2q_{\bot}^y) \de_{s_1s_3} \de_{s_2\bar{s}_4} \right)
\leea{nra18}
contributes (because of the spin structure) to the second order of BSPT:
\be
\de M^2_2 = \sum_{\mu,s_i}
 \frac{< \Phi_{100} |\de V^{(1)}| \Phi_{\mu,s_i} > \, < \Phi_{\mu,s_i} |\de V^{(1)}| \Phi_{100} >}
      {M^2_1 - M^2_n}
\; . \lee{nra19}
The order $O\left( \left(\frac{p}{m} \right)^2\right)$ 
(cf. remark after \eq{nra11}) is 
\be
\de V^{(2)} = \frac{1}{2(2\pi)^3} \frac{1}{4m} \frac{1}{2m} \,
 \left( 8e^2 \frac{q_z^2}{\vec{q}^{\,2}} \de_{s_1\bar{s}_2} \de_{s_1\bar{s}_3}
 \de_{s_2\bar{s}_4} + 8e^2 \de_{s_1s_2} \de_{s_3s_4} \de_{s_1s_3}
 +4e^2 \de_{s_1\bar{s}_3} \de_{s_2\bar{s}_4} \right)
\lee{nra20}
and contributes to the first order of BSPT:
\be
\de M^2_1 = < \Phi_{100} |\de V^{(2)}| \Phi_{100} >
\; . \lee{nra21}
Both contributions were calculated in \cite{JoPeGl} with the result
\bea
\de M^2 &=& \de M_1^2 + \de M_2^2 \nn \\
<1|\de M^2|1> &=& -\frac{5}{12} m \alpha^4 \nn \\
<2|\de M^2|2> &=& <3|\de M^2|3>= <4|\de M^2|4> = \frac{1}{6} m \alpha^4
\; , \leea{nra22}
where the eigenvectors in spin space are defined as follows:
\bea
&& |1> = \frac{1}{\sqrt{2}} \, (|\!\!+-\!\!> - |\!\!-+\!\!>) \; , \nn \\
&& |2> = \frac{1}{\sqrt{2}} \, (|\!\!+-\!\!> + |\!\!-+\!\!>) \; ,\qquad
   |3> = |\!\!--\!\!> \; ,\qquad
   |4> = |\!\!++\!\!>
\; . \leea{nra23}
Making use of the relation between Coulomb energy units
and ${\rm Ryd}=\frac{1}{2}m\alpha^2$ we have the standard result
for the singlet-triplet mass splitting for positronium,
$\frac{7}{6}\alpha^2 {\rm Ryd}$. The degeneracy of the triplet ground state
$n=1$ reflects the rotational invariance, manifest in the system in
nonrelativistic approximation.

\section{Conclusion}

Basing on the similarity scheme together with the flow equation method
we derive the renormalized to the second order $O(e^2)$ LFQED Hamiltonian.
It has, in its turn, two aspects considered further in the work.

The first point is related with renormalization group. It turnes out, that
electron (photon) mass correction and wave function renormalization 
constant vary with UV cutoff in accordance to $1$-loop renormalization group
equations. This indicates an intimate connection between Wilson's
renormalization and the flow equation method. The same equivalence
to Wilson's renormalization was shown for the projection operator techniques
(Bloch-Feshbach formalism) \cite{MuRa}.

In the case of LFQED two types of divergencies
arise: UV divergencies, associated with a large transverse momentum
and severe IR divergencies, originating   from light-front gauge singularities
and present in light-front Hamiltonian calculations for QED. Flow equation method
deliver in divergent integrals straightforwardly the regulator,
that regularize UV and partially IR divergencies. It was shown in the work,
that gauge invariant calculations of divergent terms give rise to IR finite results.
Explicitly, the IR divergencies were removed from the electron (photon) energy
correction, when all diagramms to the second order in the corresponding sectors
(arising from flow equations and normal-ordering Hamiltonian) were considered.
This enables to choose IR-independent mass counterterms and insures IR-finite 
physical masses.

To complete the renormalization group analysis the third order (coupling constant
renormalization) and Ward identities must be considered in the techniques discussed.

The second aspect is connected with the low-energy sector of the theory and 
bound state calculations. The renormalized Hamiltonian acts in the space
where the energy differences between initial and final states are restricted
by the UV cutoff, and therefore it can be interpreted as an effective Hamiltonian
descibing the low-energy physics.

Making use of diagrammatic rules for the renormalized LFQED Hamiltonian,
we obtain the electron-positron interaction to the second order.
The generated interaction, arising from elimination of $ee\ga$-vertex,
insures at any cutoff $\la$ the absence  of collinear divergencies 
in the $e\bar{e}$-interaction.

The $e\bar{e}$-interaction possesses the explicit and implicit,
through the running mass and coupling, cutoff dependence. We aimed in the work
to get rid of explicit $\la$-dependence, performing the limit $\la\rightarrow 0$
or making use of the nonrelativistic approximation, to obtain the 
physical results for the spectrum. The instant form frame was used, that manifest
several advantages as compared with the light-front one. First, the complete 
elimination of the $ee\ga$-vertex is possible, that corresponds to the limit
$\la\rightarrow 0$ where the $e\bar{e}$-interaction obviously does not
depend any more on the cutoff $\la$. In this limit the interaction is governed
by generated and instantaneous terms. Note, the light-front bound state
equation is boost and frame invariant, giving rise to the same spectrum
in both frames.

To calculate the positronium spectrum analyticaly the nonrelativistic
approximation with respect to the
electron momentum $\frac{|\vec{p}|}{m}=O(\alpha)$ and binding energy 
$\frac{B_N}{m}=O(\alpha)$ was performed. The instant form enables then to obtain
the cutoff independent $e\bar{e}$-interaction as long as the UV cutoff is restricted
to be in the nonrelativistic domain, namely $\la <\!\!< m$. The leading order
(in nonrelativistic expansion) of this interaction describes $3$d Coulomb 
interaction, giving rise to standard Coulomb energy levels and wave functions.
Note, that in the nonrelativistic approximation it occurs exact cancellation of instantaneous
interaction by the leading order generated and perturbative theory terms,
giving rise to the $3$d-Coulomb interaction. 
The next order terms define the spin structure of the $e\bar{e}$-interaction.
The same result for the spin-dependent terms was obtained in \cite{JoPeGl}, where
the ground state singlet-triplet mass splitting for positronium was calculated
to be $\frac{7}{6}\alpha^2 Ryd$. The mass degeneracy of the triplet ground state 
manifests the rotational invariance, maintained at the order $\alpha^4$.

The correct results obtained for the positronium mass spectrum in nonrelalivistic
approximation are encouraging. This makes possible to consider an exact expression
for the $e\bar{e}$-interaction, written in instant form frame, as a basis for future
numerical calculations.
 
\paragraph{Acknowledgments}

One of the authors, Elena Gubankova, would like to thank
Dr.~Brett van de Sande for very useful discussions that
improved the understanding of the problem. Also she thanks
Billy D.~Jones for discussions, held per e-mail, 
on the similarity transformation scheme; and Dr.~Stefan Kehrein
for explanation of the flow equation method.
Also E.~G. thanks Gerhard Kulzinger for many useful discussions,
constant support and help in work.
This work was partially supported by the Deutsche Forschungsgemeinschaft,
grant no. GRK 216/5-95.

\newpage
\appendix

\section{Renormalization scheme of similarity unitary transformation}

Here we discuss briefly the renormalization scheme
presented by Glazek and Wilson \cite{GlWi}, called similarity
unitary transformation, and compare it with the flow equations by Wegner.

Both approaches aim to bring the Hamiltonian to the most diagonal form,
explicitly only the matrix elements between the free states with
\be
|\Delta _{ij}| < \lambda
\lee{a1}
and $\Delta_{ij}=E_i-E_j$, are present in the renormalized Hamiltonian.
For this purposes the continuous unitary transformation must be performed
to preserve unchanged the spectrum (eigenvalues) of the initial
bare cutoff Hamiltonian. The demand of diagonal structure
does not define completly the generator of the transformation. This freedom
is used to eliminate small energy denominators in the final renormalized 
Hamiltonian. This results in a system of two self-consistent non-linear 
differential equations for the Hamiltonian $H(l)$ and the generator of the
transformation $\eta (l)$. The dependence on the continuous flow
parameter $l$ in the flow equations by Wegner is replaced by the 
cutoff dependence $\lambda$ in the similarity approach,
with the connection
\be
l=1/ \lambda ^2
\lee{a2}
in the renormalized Hamiltonians.
 
The difference of the two methods manifests the residual freedom
in the choice of the direction of the infinitesimal rotation,
actually defining how fast the non-diagonal matrix elements vanish.

We summarize the equations for both schemes, written in matrix form.
Remember, the function $f_{ij}$ defines the solution
for the leading order interaction term.

\paragraph{I.} The flow equations by Wegner \cite{We}:
\be
\frac{dH_{ij}}{dl}=[\eta,H_I]_{ij}+\frac{du_{ij}}{dl} \, \frac{H_{ij}}{u_{ij}}
\; , \lee{a3}
\be
\eta_{ij}=\frac{1}{E_i-E_j} \left(- \frac{du_{ij}}{dl} \, \frac{H_{ij}}{u_{ij}} \right)
\lee{a4}
with
\be
u_{ij} = \exp(-l \Delta_{ij}^2)
\lee{a5}
and
\be
f_{ij}=u_{ij}
\; . \lee{a6}

\paragraph{II.} The similarity unitary transformation by Gla\-zek
and Wil\-son \cite{GlWi}:
\be
\frac{dH_{ij}}{d \lambda} = u_{ij}[\eta,H_I]_{ij} +
 r_{ij} \frac{du_{ij}}{d\lambda} \, \frac{H_{ij}}{u_{ij}}
\; , \lee{a7}
\be
\eta_{ij} = \frac{r_{ij}}{E_i-E_j} \left([\eta,H_I]_{ij} -
 \frac{du_{ij}}{d \lambda} \, \frac{H_{ij}}{u_{ij}} \right)
\lee{a8}
and
\be
f_{ij}=u_{ij} \exp(r_{ij})
\; . \lee{a9}
Also the following transformation is used \cite{GlWi}:
\be
\frac{dH_{ij}}{d\lambda}=u_{ij}[\eta,H_I]_{ij}+
\frac{du_{ij}}{d\lambda} \, \frac{H_{ij}}{u_{ij}}
\; , \lee{a10}
\be
\eta _{ij} = \frac{1}{E_i-E_j} \left( r_{ij}[\eta,H_I]_{ij} -
 \frac{du_{ij}}{d \lambda} \, \frac{H_{ij}}{u_{ij}} \right)
\lee{a11}
and 
\be
f_{ij}=u_{ij}
\; , \lee{a12}
where for both similarity transformations
\be
u_{ij} = \theta (\lambda -|\Delta _{ij}|)
\lee{a13}
and
\be
u_{ij} + r_{ij} = 1
\; . \lee{a14}
Also other choices for the similarity function $u_{ij}$ with the step
behaviour are possible \cite{GlWi}. 

Remember, the function $f_{ij}$ defines the solution for the leading order
interaction term. The first order equations for $H$ and $\eta$, written
through the $f$-function are unique for both methods ({\bf I.} and {\bf II.})
\be
H_{I,ij}^{(1)}(l) = H_{I,ij}^{(1)}(l=0)
 \frac{f_{ij}(l)}{f_{ij}(l=0)}
\; , \lee{a15}
\be
\eta _{ij}^{(1)}(l)= -\frac{1}{E_i-E_j} \, \frac{dH_{I,ij}^{(1)}}{dl}
\lee{a16}
with the connection given in \eq{a2} in the renormalized values,
and $dl \rightarrow d\lambda$ implied.
This will be exploited further for the calculations in the main text.

\section{Calculation of \boldmath$[\eta^{(1)}(l), H_{ee\gamma}]$
 in the \boldmath$e\bar{e}$-sector}

Here we calculate the commutator $[\eta^{(1)}(l), H_{ee\gamma}]$
in the electron-positron sector.
The leading order generator $\eta^{(1)}$ is:
\bea
\eta^{(1)}(l) &=& \sum_{\lambda s_1s_3}\int_{p_1p_3q}\!\!\!(\eta_{p_1p_3}^*(l)
\varepsilon_{\lambda}^i\tilde{a}_q +
\eta_{p_1p_3}(l) \varepsilon_{\lambda}^{i *}\tilde{a}_{-q}^+) \,
(\tilde{b}_{p_3}^+\tilde{b}_{p_1}+\tilde{b}_{p_3}^+\tilde{d}_{-p_1}^+ +
\tilde{d}_{-p_3}\tilde{b}_{p_1}+\tilde{d}_{-p_3}\tilde{d}_{-p_1}^+)\nonumber\\
&&\hspace{4em} \times \chi_{s_3}^+ \Gamma_l^i(p_1,p_3,-q) \chi_{s_1} \,
                      \delta_{q,-(p_1-p_3)}
\; , \leea{b1}
where
\be
\eta_{p_1p_3}(l)=-\Delta_{p_1p_3} \cdot g_{p_1p_3} =
\frac{1}{\Delta_{p_1p_3}} \cdot \frac{dg_{p_1p_3}}{dl}
\; , \lee{b2}
\mbox{$\Delta_{p_1p_3}=p_1^--p_3^--(p_1-p_3)^-$},
and the electron-photon coupling
\bea
H_{ee\gamma} &=& \sum_{\lambda s_2s_4}\int_{p_2p_4q'}\!\!\!(g_{p_2p_4}^*(l)
\varepsilon _{\lambda'}^j\tilde{a}_{q'}+
g_{p_2p_4}(l)\varepsilon _{\lambda'}^{j *}\tilde{a}_{-q'}^+) \,
(\tilde{b}_{p_4}^+\tilde{b}_{p_2} +\tilde{b}_{p_4}^+\tilde{d}_{-p_2}^+ +
\tilde{d}_{-p_4}\tilde{b}_{p_2} +\tilde{d}_{-p_4}\tilde{d}_{-p_2}^+)\nonumber\\
&&\hspace{4em} \times\chi_{s_4}^+ \Gamma_l^i(p_2,p_4,-q') \chi_{s_2} \,
               \delta_{q',-(p_2-p_4)}
\; , \leea{b3}
where
\be
\Gamma_l^i(p_1,p_2,q) = 2\frac{q^i}{q^+} -
\frac{\sigma\cdot p_2^{\bot} - im}{p_2^+}\sigma^i -
\sigma^i\frac{\sigma\cdot p_1^{\bot} + im}{p_1^+}
\lee{b4}
and the tilde-fields are defined in \eq{ch19}.
Further we use the identities for the polarisation vectors and spinors
\be
\sum_{\lambda} \varepsilon_{\lambda}^{i *} \varepsilon_{\lambda}^j
= \de^{ij} \; ,\qquad
\chi_s^+ \chi_{s'} = \de_{ss'}
\; . \lee{b5}

Using the commutation relations, \eq{ch10}, and identities \eq{b5} we have 
\bea
[\eta^{(1)}(l),H_{ee\gamma}] &=& \frac{1}{2} \,
 \left( - \eta_{p_1p_3} g_{p_2p_4}^* \frac{\theta(p_1^+-p_3^+)}{p_1^+ - p_3^+}
        + \eta_{p_1p_3}^* g_{p_2p_4} \frac{\theta(p_3^+-p_1^+)}{p_3^+-p_1^+} \right) \\
&& \times {\boldmath :} \,
 (- \tilde{b}_{p_3}^+ \tilde{d}_{-p_2}^+ \tilde{d}_{-p_4} \tilde{b}_{p_1}
  - \tilde{b}_{p_4}^+ \tilde{d}_{-p_1}^+ \tilde{d}_{-p_3} \tilde{b}_{p_2}
  + \tilde{b}_{p_3}^+ \tilde{d}_{-p_1}^+ \tilde{d}_{-p_4} \tilde{b}_{p_2}
  + \tilde{b}_{p_4}^+ \tilde{d}_{-p_2}^+ \tilde{d}_{-p_3} \tilde{b}_{p_1}) \,
          {\boldmath :} \nn \\
&& \times (\chi_{s_3}^+ \Gamma_l^i(p_1,p_3,p_1-p_3) \chi_{s_1}) \,
          (\chi_{s_4}^+ \Gamma_l^i(p_2,p_4,p_2-p_4) \chi_{s_2}) \:
          \delta_{p_1+p_2,p_3+p_4} \nn
\; , \leea{b6}
where the first two terms of the field operators contribute to
the exchange channel, and the next two to the annihilation channel.
We take into account both $s$- and $t$-channel terms to calculate
the bound states. The $:\;:$ stand for the normal ordering of the
fermion operators and $(\frac{1}{2})$ is the symmetry factor. The
sum over helicities $s_i$ and the 3-dimensional integration 
over momenta $p_i$, $i=1,..4$, according to \eq{ch20} is implied. 
We rewrite for both channels
\be
\hspace{0cm}
[\eta,H_{ee\gamma}]=
\left\{ \begin{array}{l}
  M_{2ij}^{(ex)}(\frac{1}{2})
       \left\{ \frac{\theta(p_1^+-p_3^+)}{(p_1^+-p_3^+)} \right.
       (\eta_{p_1,p_3}g_{-p_4,-p_2}^*-\eta_{-p_4,-p_2}^*g_{p_1,p_3}) \\
\hspace{2.0cm}
     + \left. \frac{\theta(-(p_1^+-p_3^+))}{-(p_1^+-p_3^+)}
       (\eta_{-p_4,-p_2}g_{p_1,p_3}^*-\eta_{p_1,p_3}^*g_{-p_4,-p_2})\ \right\} \\
\hspace{1.8cm}
     \times \delta^{ij}\delta_{p_1+p_2,p_3+p_4} \,
     b_{p_3s_3}^+d_{p_4\bar{s}_4}^+d_{p_2\bar{s}_2}b_{p_1s_1} \\
\\
 -M_{2ij}^{(an)}(\frac{1}{2})
       \left\{ \frac{\theta(p_1^++p_2^+)}{(p_1^++p_2^+)} \right.
       (\eta_{p_1,-p_2}g_{-p_4,p_3}^*-\eta_{-p_4,p_3}^*g_{p_1,-p_2}) \\
\hspace{2.2cm}
       + \left. \frac{\theta(-(p_1^++p_2^+))}{-(p_1^++p_2^+)}
       (\eta_{-p_4,p_3}g_{p_1,-p_2}^*-\eta_{p_1,-p_2}^*g_{-p_4,p_3}) \right\} \\
\hspace{2.0cm}
       \times \delta^{ij}\delta_{p_1+p_2,p_3+p_4} \,
       b_{p_3s_3}^+d_{p_4\bar{s}_4}^+d_{p_2\bar{s}_2}b_{p_1s_1}
\end{array} \right. 
\lee{b7}
where 
\bea
M_{2ij}^{(ex)}
&=& (\chi_{s_3}^+ \Gamma_l^i(p_1,p_3,p_1-p_3) \chi_{s_1}) \,
    (\chi_{\bar{s}_2}^+ \Gamma_l^j(-p_4,-p_2,-(p_1-p_3)) \chi_{\bar{s}_4}) \nn \\
\\
M_{2ij}^{(an)}
&=& (\chi_{s_3}^+ \Gamma_l^i(-p_4,p_3,-(p_1+p_2)) \chi_{\bar{s}_4}) \,
    (\chi_{\bar{s}_2}^+ \Gamma_l^j(p_1,-p_2,p_1+p_2) \chi_{s_1}) \nn
\; . \leea{b8}

The first term in the exchange channel with $p_1^+ > p_3^+$ corresponds to the 
light-front time ordering $x_1^+ < x_3^+$ 
with the intermediate state $P_k^-=p_3^- + (p_1-p_3)^- + p_2^-$,
the second term $p_1^+<p_3^+$ and $x_1^+>x_3^+$
has the intermediate state $P_k^- = p_1^- - (p_1 - p_3)^- + p_4^-$.
Both terms can be viewed as the retarded photon exchange.
The same does hold for the annihilation channel.

Consider only real couplings and take into account the symmetry
\be
\eta_{-p_4,-p_2}=-\eta_{p_4,p_2} \; ,\qquad g_{-p_4,-p_2}=g_{p_4,p_2}
\; . \lee{b9}
Then \mbox{$\left. \left< \! p_3s_3,p_4\bar{s}_4 \right|
[\eta^{(1)},H_{ee\gamma}] \left| p_1s_1,p_2\bar{s}_2 \right> \right.$},
the matrix element of the commutator between the free states of positronium
in the exchange and annihilation channel, reads
\be
<[\eta^{(1)},H_{ee\gamma}]>/\delta_{p_1+p_2,p_3+p_4}= 
\left \{\begin{array}{l}
 M_{2ii}^{ex} \, \frac{1}{(p_1^+-p_3^+)} \,
  (\eta_{p_1,p_3}g_{p_4,p_2} + \eta_{p_4,p_2}g_{p_1,p_3})
\\
-M_{2ii}^{an} \, \frac{1}{(p_1^++p_2^+)} \,
  (\eta_{p_1,-p_2}g_{p_4,-p_3} + \eta_{p_4,-p_3}g_{p_1,-p_2})
\end{array} \right. 
\; . \lee{b10}

We rewrite this expression through the corresponding $f$-functions
\bea
\eta_{p_1,p_3} g_{p_4,p_2} + \eta_{p_4,p_2} g_{p_1,p_3}
&=& e^2 \left[
  \frac{1}{\Delta_{p_1,p_3}} \frac{df_{p_1,p_3}(l)}{dl} f_{p_4,p_2}(l)
+ \frac{1}{\Delta_{p_4,p_2}} \frac{df_{p_4,p_2}(l)}{dl}f_{p_1,p_3}(l) \right] \nn \\
\\
\eta_{p_1,-p_2} g_{p_4,-p_3} +\eta_{p_4,-p_3} g_{p_1,-p_2}
&=& e^2 \left[
  \frac{1}{\Delta_{p_1,-p_2}}\frac{df_{p_1,-p_2}(l)}{dl}f_{p_4,-p_3}(l)
+ \frac{1}{\Delta_{p_4,-p_3}}\frac{df_{p_4,-p_3}(l)}{dl}f_{p_1,-p_2}(l) \right] \nn
\leea{b11}
with $\Delta_{p_1,p_2}=p_1^--p_2^--(p_1-p_2)^-$. As we have mentioned
in Appendix A this form in terms of the $f$-function is universal
for all unitary transformations. We exploit further this expression by
specifying the $f$-function to compare the effective interactions
in different renormalization schemes (see Appendix C).    

We calculate the matrix elements \mbox{$M_{2ii}$}, eq.~(188), for both channels.
Here we follow the notations introduced in \cite{JoPeGl}.

We make use of the identities
\be
\chi_s^+\sigma^i\sigma^j \chi_s = \delta^{ij}+is\varepsilon^{ij} \; ,\qquad
\chi_s^+\sigma^j\sigma^i \chi_s = \delta^{ij}+i\bar{s}\varepsilon^{ij}
\lee{b12}
with $\bar{s} = -s$ and $\chi_s^+ \chi_{s'} = \de_{ss'}$; also of
\be
\chi_{\bar{s}}^+ \sigma^i\chi_s = -\sqrt{2}s \varepsilon_s^i \; ,\qquad
\chi_s^+\sigma^i \chi_{\bar{s}} = -\sqrt{2}s \varepsilon_s^{i*}
\lee{b13}
with \mbox{$\varepsilon_s^* = -\varepsilon_{\bar{s}}$} and
\mbox{$\varepsilon_s^i \varepsilon_{s'}^i = -\delta_{s\bar{s'}}$}.

We use the standard light-front frame, \fig{reneebarint},
\bea
&& p_1 = (xP^+,xP^{\bot}+k_\bot) \; , \hspace{3em}
   p_2 = ((1-x)P^+,(1-x)P^{\bot}-k_\bot) \; , \nn \\
&& p_3 = (x'P^+,x'P^{\bot}+k'_\bot) \; , \hspace{1.5em}
   p_4 = ((1-x')P^+,(1-x')P^{\bot}-k'_\bot)
\; , \leea{b14}
where \mbox{$P=(P^+,P^{\bot})$} is the positronium momentum.

Then, to calculate the matrix element $M_{2ii}$ in the {\bf exchange channel}, we find
\bea
P^+[\chi_{s_3}^+\Gamma^i(p_1,p_3,p_1-p_3)\chi_{s_1}]
&=& \chi_{s_3}^+ \left[ 2\frac{(k_\bot-k'_\bot)^i}{(x-x')}
 - \frac{\sigma \cdot k'_\bot}{x'} \sigma^i+\sigma^i\frac{\sigma \cdot k_\bot}{x} +
          im\frac{x-x'}{xx'}\sigma^i
   \right] \chi_{s_1} \nn \\
&=& T_2^i \delta_{s_1s_3} +
 im\frac{x-x'}{xx'} (-\sqrt{2}) s_1 \varepsilon_{s_1}^i \delta_{s_1\bar{s}_3}
\; , \leea{b15}
and
\bea
&&\hspace{-1cm}
P^+[\chi_{\bar{s}_2}^+ \Gamma^i(-p_4,-p_2,-(p_4-p_2)) \chi_{\bar{s}_4}] \nn \\
&&\hspace{2cm} = \chi_{\bar{s}_2}^+
 \left[ 2 \frac{(k_\bot-k'_\bot)^i}{x-x'}
 + \left( \frac{\sigma \cdot k_\bot}{1-x} \sigma^i
           + \sigma^i \frac{\sigma \cdot k'_\bot}{1-x'} \right)
 - im \frac{x-x'}{(1-x)(1-x')} \sigma^i
 \right] \chi_{\bar{s}_4} \nn \\
&&\hspace{2cm} = - \left[ T_1^i\delta_{s_2s_4}
 + im \frac{x-x'}{(1-x)(1-x')} (-\sqrt{2}) s_2 \varepsilon_{s_2}^i
 \delta_{s_2\bar{s}_4} \right]
\; , \leea{b16}
where we have introduced
\bea
T_1^i & \equiv& -\left[ 2 \frac{(k_\bot-k'_\bot)^i}{x-x'}+\frac{k_\bot^i(s_2)}{(1-x)} +
 \frac{{k'}_\bot^i(\bar{s}_2)}{(1-x')} \right] \nn \\
\\
T_2^i & \equiv& 2 \frac{(k_\bot-k'_\bot)^i}{x-x'}-\frac{k_\bot^i(s_1)}{x} -
 \frac{{k'}_\bot^i(\bar{s}_1)}{x'} \nn
\leea{b17}
and
\be
k_\bot^i(s)\equiv k_\bot^i+is \, \varepsilon_{ij} \, k_\bot^j
\; . \lee{b18}
Finaly we result 
\bea
P^{+2} \, M_{2ii}^{(ex)} &\hspace{-2mm}=\hspace{-2mm}& - \left\{
   \delta_{s_1s_3} \delta_{s_2s_4} T_1^{\bot} \cdot T_2^{\bot}
 - \delta_{s_1\bar{s}_2} \delta_{s_1\bar{s}_3} \delta_{s_2\bar{s}_4}
 2m^2 \frac{(x-x')^2}{xx'(1-x)(1-x')} \right. \\
&&\hspace{2em} \left.
 + im \sqrt{2}(x'-x) \left[ \delta_{s_1\bar{s}_3} \delta_{s_2s_4}
 \frac{s_1}{xx'}T_1^{\bot} \cdot \varepsilon_{s_1}^{\bot}
 + \delta_{s_1s_3} \delta_{s_2\bar{s}_4}
 \frac{s_2}{(1-x)(1-x')} T_2^{\bot} \cdot \varepsilon_{s_2}^{\bot} \right]
 \right\} \nn
\; . \leea{b19}
Whereas in the {\bf annihilation channel} we calculate
\bea
P^+[\chi_{s_3}^+ \Gamma^i(-p_4,p_3,-(p_1+p_2)) \chi_{\bar{s}_4}]
&=& \chi_{s_3}^+ \left[
 - \frac{\sigma \cdot k'_\bot}{x'} \sigma^i + \sigma^i \frac{\sigma \cdot k'_\bot}{1-x'}
 + im\frac{1}{x'(1-x')}\sigma^i \right] \chi_{\bar{s}_4} \nn \\
&=& T_3^i\delta_{s_3\bar{s}_4}
 + im\frac{1}{x'(1-x')} (-\sqrt{2}) s_4 \varepsilon_{s_4}^{i*}
 \delta_{s_3s_4}
\leea{b20}
and
\bea
\hspace{-1cm}
P^+[\chi_{\bar{s}_2}^+ \Gamma^i(p_1,-p_2,p_1+p_2) \chi_{s_1}]
&=& \chi_{\bar{s}_2}^+ \left[
 \frac{\sigma \cdot k_\bot}{1-x} \sigma^i - \sigma^i \frac{\sigma \cdot k_\bot}{x}
 - im\frac{1}{x(1-x)} \sigma^i \right] \chi_{s_1} \nn \\
&=& T_4^i\delta_{s_1\bar{s}_2}
 -im\frac{1}{x(1-x)} (-\sqrt{2}) s_1 \varepsilon_{s_1}^i
 \delta_{s_1s_2}
\; , \leea{b21}
where we have introduced
\bea
T_3^i & \equiv & -\frac{{k'}_\bot^i(\bar{s}_3)}{x'}
 + \frac{{k'}_\bot^i(s_3)}{1-x'} \nn \\
\\
T_4^i & \equiv & \frac{k_\bot^i(\bar{s}_1)}{1-x}
 - \frac{k_\bot^i(s_1)}{x} \nn
\; . \leea{b22}
We finally have
\bea
P^{+2} \, M_{2ii}^{(an)} &=& \delta_{s_1\bar{s}_2} \delta_{s_3\bar{s}_4}
 T_3^{\bot}\cdot T_4^{\bot}
 + \delta_{s_1s_2} \delta_{s_3s_4}\delta_{s_1s_3}
 2m^2 \frac{1}{xx'(1-x)(1-x')} \\
&& + im \sqrt{2} \left[\delta_{s_3\bar{s}_4}\delta_{s_1s_2}
 \frac{s_1}{x(1-x)} T_3^{\bot} \cdot \varepsilon_{s_1}^{\bot}
 - \delta_{s_3s_4} \delta_{s_1\bar{s}_2}
 \frac{s_3}{x'(1-x')} T_4^{\bot}\cdot \varepsilon_{s_4}^{\bot *} \right] \nn
\; . \leea{b23}

\section{Generated interaction in comparison with the result derived in the
 renormalization scheme of Gla\-zek and Wil\-son}

We derive here the generated interaction to the order $O(e^2)$
in the renormalization scheme of Gla\-zek and Wil\-son and compare it
with the one derived in the scheme of Wegner. To this end we use
the flow equations of type {\bf II.}, see Appendix A,
with, for sake of simplicity, the $f$-function chosen as
\be
f_{p_ip_f,\lambda}= \theta(\lambda-|\Delta_{p_ip_f}|)
\; , \lee{c1}
where $\Delta_{p_ip_f}= \sum p_i^--\sum p_f^-$. This gives 
for the second order generated interaction 
in $|e\bar{e}>$-sector
\be
\frac{dV_{p_ip_f,\lambda}^{gen}}{d\lambda}\\
=f_{p_ip_f,\lambda}<[\eta^{(1)}_{\lambda},H_{\lambda}^{ee\gamma}]>_{|e\bar{e}>}
+\frac{df_{p_ip_f,\lambda}}{d\lambda}
\frac{V_{p_ip_f,\lambda}^{eff}}{f_{p_ip_f,\lambda}}
\lee{c2}
with the solution 
\be
V_{p_ip_f,\lambda}=-f_{p_ip_f,\lambda}\int_{\lambda}^{\infty}\\
d\lambda'<[\eta^{(1)}_{\lambda'},H_{\lambda'}^{ee\gamma}]>_{|e\bar{e}>}
\; , \lee{c3}
where the initial condition
$V_{p_ip_f}(\lambda\rightarrow\infty)=0$ is implied.
We make use of the expression for the matrix element
of the commutator $[\eta ,H^{ee\gamma}]$
in the exchange and annihilation channels, eqs.~(190) and (191) in Appendix B,
where now the derivative is performed
with respect of the UV cutoff, i.e. $d/d\lambda$. This results
for the generated interaction in both channels
\bea
V_{\lambda}^{(ex)} &=& -e^2M_{2ii,\lambda}^{(ex)} \, \frac{1}{p_1^+-p_3^+} \,
f_{p_ip_f,\lambda}
\left[
 \frac{\int_{\lambda}^{\infty}\frac{df_{p_1,p_3,\lambda'}}{d\lambda'}
 f_{p_4,p_2,\lambda'}d\lambda'}{\Delta_{p_1,p_3,\lambda}}+
 \frac{\int_{\lambda}^{\infty}\frac{df_{p_4,p_2,\lambda'}}{d\lambda'}
 f_{p_1,p_3,\lambda'}d\lambda'}{\Delta_{p_4,p_2,\lambda}}
\right] \nn \\
\\
V_{\lambda}^{(an)} &=& e^2M_{2ii,\lambda}^{(an)} \, \frac{1}{p_1^++p_2^+} \,
f_{p_ip_f,\lambda}
\left[
 \frac{\int_{\lambda}^{\infty}\frac{df_{p_1,-p_2,\lambda'}}{d\lambda'}
 f_{p_4,-p_3,\lambda'}d\lambda'}{\Delta_{p_1,-p_2,\lambda}}+
 \frac{\int_{\lambda}^{\infty}\frac{df_{p_4,-p_3,\lambda'}}{d\lambda'}
 f_{p_1,-p_2,\lambda'}d\lambda'}{\Delta_{p_4,-p_3,\lambda}}
\right] \nn
\leea{c4}
and with $f_{p_1p_2}$ being according to \eqs{a12}{a13}
\bea
V_{\lambda}^{(ex)} &\hspace{-2mm}=\hspace{-2mm}&
-e^2M_{2ii,\lambda}^{(ex)} \, \frac{1}{p_1^+-p_3^+} \,
 \theta(\lambda - |\Delta_{p_ip_f,\lambda}|) \nn \\
&&\hspace{-2mm} \times \left[ \frac{\theta(|\Delta_{p_1p_3}|-|\Delta_{p_4p_2}|)
   \theta(|\Delta_{p_1p_3}|-\lambda)}{\Delta_{p_1p_3}}
   +\frac{\theta(|\Delta_{p_4p_2}|-|\Delta_{p_1p_3}|)
   \theta(|\Delta_{p_4p_2}|-\lambda)}{\Delta_{p_4p_2}}
 \right] \nn \\
\\
V_{\lambda}^{(an)} &\hspace{-2mm}=\hspace{-2mm}&
e^2M_{2ii,\lambda}^{(an)} \, \frac{1}{p_1^++p_2^+} \,
 \theta(\lambda - |\Delta_{p_ip_f,\lambda}|) \nn \\
&&\hspace{-2mm} \times \left[ \frac{\theta(|\Delta_{p_1,-p_2}|-|\Delta_{p_4,-p_3}|)
   \theta(|\Delta_{p_1,-p_2}|-\lambda)}{\Delta_{p_1,-p_2}}
   + \frac{\theta(|\Delta_{p_4,-p_3}|-|\Delta_{p_1,-p_2}|)
   \theta(|\Delta_{p_4,-p_3}|-\lambda)}{\Delta_{p_4,-p_3}}
 \right] \nn
\; . \leea{c5}

This result for the generated $e\bar{e}$-interaction was obtained by the authors
of \cite{JoPeGl}. It is to be compared with the expression in the main text
for the generated interaction, eq.~(50). First, due to the $\theta$-functions
the interaction in eq.~(208) is also free of divergencies coming from
small energy denominators. As compared with the
result of flow equations, eq.~(50), each energy denominator in eq.~(208) has its
own relative weight in the scheme of similarity transformation.
We rewrite eq.~(208) as the sum of two terms, the first corresponding
to the result of flow equations, eq.~(208), and a second term representing the rest,
which carries different weights for each energy denominator.

\bea
V_{\lambda}^{(ex)} &=& -e^2M_{2ii,\lambda}^{(ex)} \, \frac{1}{p_1^+-p_3^+} \,
\theta(\lambda-| \Delta_{p_ip_f,\lambda}|) \nn \\
&&\hspace{0em}
\times \left\{
\frac{1}{2} \left( \frac{1}{\Delta_{p_1,p_3}}+\frac{1}{\Delta_{p_4,p_2}} \right) \:
  \biggl[ 1 - \th(\la-|\De_{p_1,p_3}|) \, \th(\la-|\De_{p_4,p_2}|) \biggr]
\right. \nn \\
&&\hspace{1em}
+ \frac{1}{2} \left( \frac{1}{\De_{p_1,p_3}}-\frac{1}{\Delta_{p_4,p_2}} \right) \:
\biggl[ \theta(|\Delta_{p_1,p_3}|-|\Delta_{p_4,p_2}|) \,
        \theta(|\Delta_{p_1,p_3}|-\lambda) \nn \\
&&\hspace{10.8em}
      - \theta(|\Delta_{p_4,p_2}|-|\Delta_{p_1,p_3}|) \,
        \theta(|\Delta_{p_4,p_2}|-\lambda) \biggr]
\Biggr\} \nn \\ 
V_{\lambda}^{(an)} &=& e^2M_{2ii,\lambda}^{(an)} \, \frac{1}{p_1^++p_2^+} \,
\theta(\lambda-|\Delta_{p_ip_f,\lambda}|) \nn \\
&&\hspace{0em}
\times \left\{
\frac{1}{2} \left( \frac{1}{\Delta_{p_1,-p_2}} + \frac{1}{\Delta_{p_4,-p_3}} \right) \:
  \biggl[ 1 - \th(\la-|\De_{p_1,-p_2}|) \, \th(\la-|\De_{p_4,-p_3}|) \biggr]
\right. \nn \\
&&\hspace{1em}
+\frac{1}{2} \left( \frac{1}{\De_{p_1,-p_2}} - \frac{1}{\Delta_{p_4,-p_3}} \right) \:
\biggl[ \theta(|\Delta_{p_1,-p_2}|-|\Delta_{p_4,-p_3}|) \,
        \theta(|\Delta_{p_1,-p_2}|-\lambda) \nn \\
&&\hspace{12em}
      - \theta(|\Delta_{p_4,-p_3}|-|\Delta_{p_1,-p_2}|) \,
        \theta(|\Delta_{p_4,-p_3}|-\lambda) \biggr]
\Biggr\} 
\; . \leea{c6}
Here we have used the identity
\be
\frac{\th_1}{\De_1} + \frac{\th_2}{\De_2} =
\frac{1}{2}(\th_1+\th_2) \left( \frac{1}{\De_1}+\frac{1}{\De_2} \right)
+ \frac{1}{2}(\th_1-\th_2) \left( \frac{1}{\De_1} - \frac{1}{\De_2} \right)
\; , \lee{c7}
where in the exchange channel
\bea
\th_1 &=& \int_{\lambda}^{\infty}\frac{df_{p_1,p_3,\lambda'}}{d\lambda'}
f_{p_4,p_2,\lambda'}d\lambda'\nonumber\\
\\
\th_2 &=& \int_{\lambda}^{\infty}\frac{df_{p_1,-p_2,\lambda'}}{d\lambda'}
f_{p_4,-p_3,\lambda'}d\lambda'\nonumber
\leea{c8}
and
\be
\th_1+\th_2 = 1 - f_{p_1p_3}f_{p_4p_2}
\; . \lee{c9}
The term corresponding to the annihilation channel is treated along the same line.

For completeness, we rewrite the result
produced by the flow equations of Wegner, eq.~(50), as follows
\bea
V_{gen}^{(ex)}(l_\la) &=& -e^2M_{2ii}^{(ex)} \, \frac{1}{p_1^+-p_3^+} \,
\frac{1}{2} \left( \frac{1}{\Delta_{p_1,p_3}} + \frac{1}{\Delta_{p_4,p_2}} \right) \,
\left( 1 - {\rm e}^{-2\frac{\Delta_{p_1,p_3}}{\la} \frac{\Delta_{p_4,p_2}}{\la}} \right) \cdot
{\rm e}^{- \left( \frac{\Delta_{p_ip_f}}{\la} \right) ^2} \nn \\
\\
V_{gen}^{(an)}(l_\la) &=& e^2M_{2ii}^{(an)} \, \frac{1}{p_1^++p_2^+} \,
\frac{1}{2} \left( \frac{1}{\Delta_{p_1,-p_2}} + \frac{1}{\Delta_{p_4,-p_3}} \right)
\left( 1 - {\rm e}^{-2\frac{\Delta_{p_1,-p_2}}{\la} \frac{\Delta_{p_4,-p_3}}{\la}} \right) \cdot
{\rm e}^{- \left( \frac{\Delta_{p_ip_f}}{\la} \right)^2} \nn
\; . \leea{c10}

This interaction corresponds
to the first term in eq.~(208). In the nonrelativistic approximation
the second term with the difference of energy denominators
vanishes; hence it does not contribute to the spectrum
of positronium (see main text). In general it is an open question,
whether the second term contributes to the physical values.
This would mean the two methods not to be equivalent in general.

\section{Fermion and photon self energy terms}
 
We calculate here the fermion and photon self energy terms,
arising from the second order commutator $[\eta^{(1)},H_{ee\ga}]$.

\begin{figure}
\setlength{\unitlength}{1mm}
\begin{picture}(170,34.61)
\put(0,0){\makebox(56,34.61){ \loadeps{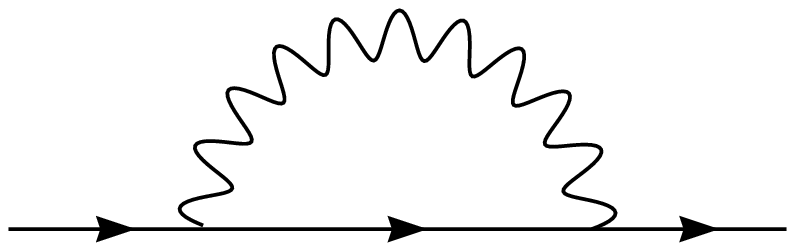} }}
\put(10,14){$p$}
\put(42,14){$p$}
\put(26,32){$k$}
  \put(57,0){\makebox(56,34.61){ \loadeps{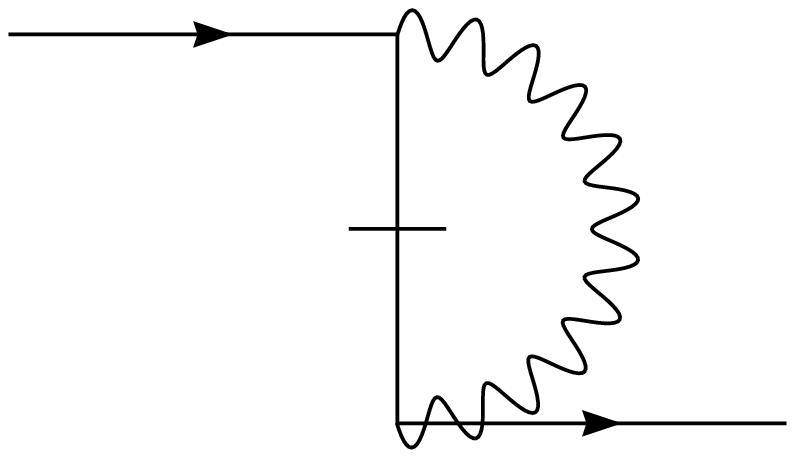} }}
  \put(67,30){$p$}
  \put(100,16){$k$}
  \put(100,3){$p$}
    \put(114,0){\makebox(56,34.61){ \loadeps{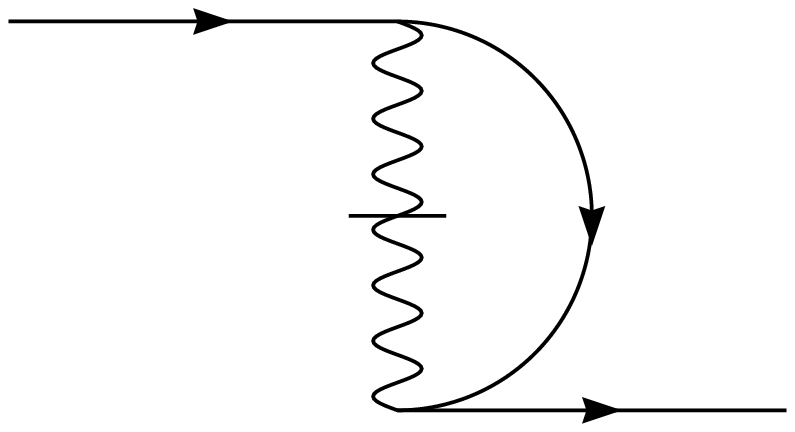} }}
    \put(124,30){$p$}
    \put(155,16){$k$}
    \put(157,3){$p$}
\put(55,17.305){$+$}
\put(112,17.305){$+$}
\end{picture}
\caption{Electron self energy: the first diagram corresponds to
the commutator term $[\eta^{(1)},H_{ee\ga}]$ in the electron self energy sector, 
next two diagrams arise from the normal ordering of instantaneous interactions.}
\label{eselfen}
\end{figure}

\paragraph{I.}
We first derive the {\bf electron self energy} terms.
Making use of the expressions for the generator of the unitary
transformation $\eta^{(1)}$ defined in \eq{gi1} and of $H_{ee\ga}$, \eq{ch14},
we obtain the following expression for the commutator in the
electron self energy sector
\bea
&&\hspace{-5em} \frac{1}{2} (\et_{p_1p_2}g_{p_2p_1}-\et_{p_2p_1}g_{p_1p_2}) \,
\biggl[ \th(p_1^+) \frac{\th(p_2^+-p_1^+)}{p_2^+ - p_1^+} \th(p^+_2) \,
               b^+_{p_2}b_{p_2}\chi^+_{s_2}\chi_{s_2}\nn\\
&&\hspace{6em}
      - \th(p_2^+) \frac{\th(p_1^+-p_2^+)}{p_1^+ - p_2^+} \th(p^+_1) \,
               b^+_{p_1}b_{p_1}\chi^+_{s_1}\chi_{s_1} \biggr]
M_{2ij}(p_1,p_2) \de^{ij}
\; , \leea{d1}
where
\be
M_{2ij}(p_1,p_2)=\Ga^i(p_1,p_2,p_1-p_2)\Ga^j(p_2,p_1,p_2-p_1)
\lee {d2}
and the momentum integration over $p_1,p_2$ is implied;
$1/2$ stands as the symmetry factor.
The matrix element of the commutator between the free fermion states is
\bea
&& <p_1,s_1|[\et^{(1)},H_{ee\ga}]|p_1,s_1>_{self energy} \nn \\
&&\hspace{5em} = -\int_{p_2}(\et_{p_1p_2} g_{p_2p_1} - \et_{p_2p_1}g_{p_1p_2}) \,
\th(p_2^+) \frac{\th(p_1^+-p_2^+)}{p_1^+-p_2^+} \,M_{2ii}(p_1,p_2)
\; , \leea{d3}
where the integration $\int_p$ is defined in \eq{ch20}. We use the expression
for the generator $\eta$ through the coupling, namely 
\be
\et_{p_1p_2} g_{p_2p_1} - \et_{p_2p_1} g_{p_1p_2} = \frac{1}{\De_{p_1p_2}}
\left( g_{p_1p_2} \frac{dg_{p_2p_1}}{dl} +
       g_{p_2p_1} \frac{dg_{p_1p_2}}{dl} \right)
\; . \lee{d4}
Change of the variables according to
\bea
p_1 &=& p \nn \\
p_2 &=& p_k \nn \\
p_1 - p_2 &=& k
\leea{d5}
brings the integral in \eq{d3} to the standard form of loop integration 
\be
-\int_k(\et_{p, p - k} g_{p - k, p} - \et_{p - k, p} g_{p, p - k}) \,
\th(p^+-k^+) \frac{\th(k^+)}{k^+} \, M_{2ii}(p,p-k)
\; . \lee{d6}
According to \eq{ri2}, the integral $\int_{l_{\la}}^{l_{\La}}$
of the commutator $[\eta^{(1)},H_{ee\ga}]$ defines the difference between
the energies (or energy corrections) \mbox{$\de p_{1\la}^--\de p_{1\La}^-$}.
Making use of
\be 
\int^{l_{\la}}_{l_{\La}} dl' (\et_{p_1p_2} g_{p_2p_1}-\et_{p_2p_1} g_{p_1p_2})
= \frac{1}{p_1^- - p_2^- - (p_1-p_2)^-} \,
(g_{p_1, p_2, \La} g_{p_2, p_1, \la} - g_{p_1, p_2, \la} g_{p_2, p_1, \La})
\lee{d7}
we have the following explicit expression: 
\bea 
&&\hspace{-1em} \de p_{1\la}^--\de p_{1\La}^- = e^2\int \frac{d^2k^{\bot}dk^+}{2(2\pi)^3} \,
\frac{\th (k^+)}{k^+}
\th (p^+-k^+) \, \frac{(-1)}{p^--k^--(p-k)^-} \\
&&\hspace{4em} \times \Ga^i(p-k,p,-k) \Ga^i(p,p-k,k) \,
\left[ \exp\left\{-2 \left( \frac{\De_{p,p-k}}{\la} \right)^2
\right\}
- \exp\left\{-2 \left( \frac{\De_{p,p-k}}{\La} \right)^2 \right\} \right] \nn
\; , \leea{d8}
where the solution for the $ee\ga$-coupling constant was used.
Therefore the electron energy correction corresponding to the first diagram,
cf. \fig{eselfen}, is 
\bea
\de p_{1\la}^- &=& e^2\int \frac{d^2k^{\bot}dk^+}{2(2\pi)^3} \,
 \frac{\th (k^+)}{k^+} \th (p^+-k^+) \\
&&\hspace{2em} \times \Ga^i(p-k,p,-k) \Ga^i(p,p-k,k) \,
 \frac{1}{p^--k^--(p-k)^-} \, \times (-R) \nn
\; , \leea{d9}
where we have introduced the regulator $R$, defining the cutoff condition
(see main text),
\be
R = \exp\left\{-2 \left( \frac{\De_{p,k}}{\la} \right)^2 \right\}
\lee{d10}
(note that $\De_{p,k}=\De_{p,p-k}$).
To perform the integration over $k=(k^+,k^{\bot})$ explicitly, choose
the parametrization
\bea
\frac{k^+}{p^+} &=& x \nn \\
k &=& (xp^+,xp^{\bot}+\kappa^{\bot})
\; , \leea{d11}
where $p=(p^+,p^{\bot})$ is the external electron momentum.
Then the terms occuring in $\de p_{1\la}^-$ are rewritten in the form
\bea
& \Ga^i(p-k,p,-k)\Ga^i(p,p-k,k)=
\frac{1}{(p^+)^2(1-x)^2} \left( \left( 4\frac{1}{x^2}-4\frac{1}{x}+2 \right)
\kappa_{\bot}^2+2m^2x^2 \right) & \nn \\
& \De_{p, p - k} = p^- - k^- - (p - k)^- = \frac{1}{p^+x(1-x)}
(x(1-x)p^2 - \kappa_{\bot}^2-xm^2) = \frac{\tilde{\De}_{p,p-k}}{p^+} &
\; . \leea{d12}
Therefore the integral for the electron energy correction
corresponding to the first diagram of \fig{eselfen} takes the form
\bea
p^+\de p_{1\la}^- &=&-\frac{e^2}{8\pi^2}\int_0^1dx\int d\kappa_{\bot}^2
\frac{(\frac{2}{x^2}-\frac{2}{x}+1)\kappa_{\bot}^2+m^2x^2}
{(1-x)(\kappa_{\bot}^2+f(x))}\times(-R) \\
&=& -\frac{e^2}{8\pi^2}\int_0^1dx\int d\kappa_{\bot}^2 \nn \\
&&\hspace{2em} \times \left[ \frac{p^2-m^2}{\kappa_{\bot}^2+f(x)}
 \left( \frac{2}{[x]}-2+x \right) - \frac{2m^2}{\kappa_{\bot}^2+f(x)}
 + \left( \frac{2}{[x]^2}+\frac{1}{[1-x]} \right)
\right] \times (-R) \nn
\; , \leea{d13}
where 
\be
f(x)=xm^2 - x(1-x)p^2
\; . \lee{d14}
In the last integral the principal value prescription  for $\frac{1}{[x]}$
as $x\rightarrow 0$ was introduced (see main text), to regularize
the IR divergencies present in the longitudinal direction. 

We thus have derived the expression for the energy correction
which has been used in the main text.

\begin{figure}
\setlength{\unitlength}{1mm}
\begin{picture}(170,34.61)
\put(28,0){\makebox(56,34.61){ \loadeps{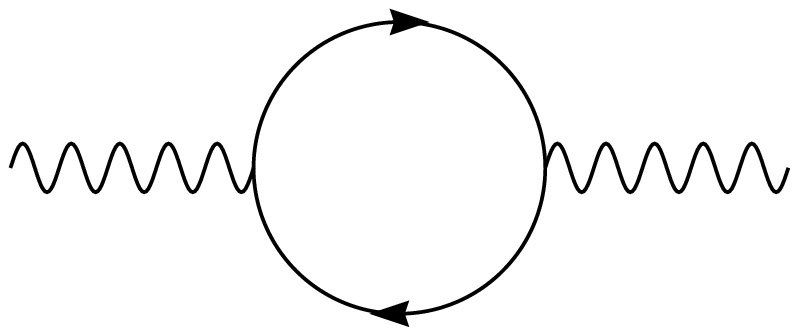} }}
\put(38,12){$p$}
\put(70,12){$p$}
\put(54,28){$k$}
  \put(86,0){\makebox(56,34.61){ \loadeps{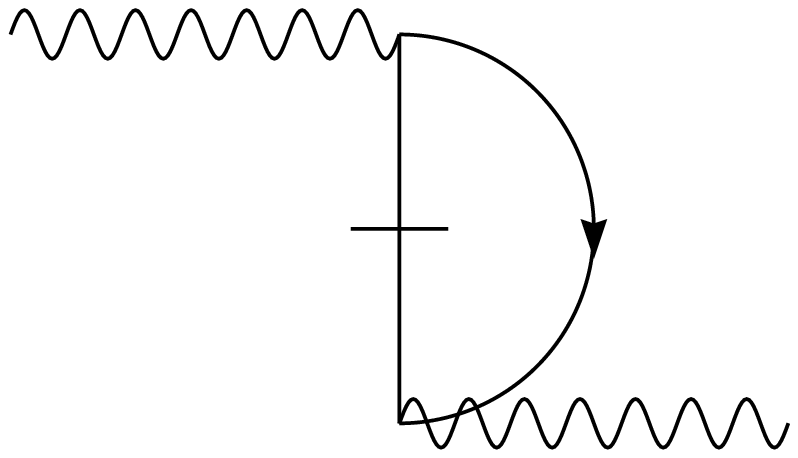} }}
  \put(96,31){$p$}
  \put(126,16){$k$}
  \put(129,3){$p$}
\put(84,17.305){$+$}
\end{picture}
\caption{Photon self energy: the first diagram comes from the commutator
$[\eta^{(1)},H_{ee\ga}]$ in the photon self energy sector, the second one
from the normal ordering of the instantaneous interaction.}
\label{photselfen}
\end{figure}

\paragraph{II.}
We repeat the same procedure for the {\bf photon self energy}.
The second order commutator $[\eta^{(1)},H_{ee\ga}]$ gives
the following expression in the photon self energy sector
\bea
&&\hspace{-2em} \frac{1}{2} (\et_{p_1p_2} g_{p_2p_1} - 
\et_{p_2p_1} g_{p_1p_2}) \cdot
\biggl[ \th(p_1^+) \th(-p_2^+) \frac{\th(p_1^+-p_2^+)}{(p_1^+-p_2^+)} \,
            a_{-q}^+ a_{-q} \varep_{\la}^{i*} \varep_{\la}^j \\
&&\hspace{9em}
      - \th(-p_1^+) \th(p_2^+) \frac{\th(p_2^+-p_1^+)}{(p_2^+-p_1^+)} \,
            a_{q}^+a_{q} \varep_{\la}^{i} \varep_{\la}^{j*} \biggr] \cdot
Tr M_{2ij}(p_1,p_2) \, \de_{q,-(p_1 - p_2)} \nn
\; , \leea{d15}
where $M_{2ij}(p_1,p_2)$ is defined in \eq{d2} and the trace acts in spin space;
the integration over the momenta $q$, $p_1$ and $p_2$ is implied.
The matrix element between the free photon states reads
\bea
&&\hspace{-2em} <q,\la|[\et^{(1)},H_{ee\ga}]|q,\la>_{self energy}\de_{ij} \\
&&\hspace{3em}
= - \frac{1}{q^+} \int_{p_1,p_2} (\et_{p_1p_2}g_{p_2p_1}-
\et{p_2p_1}g_{p_1p_2}) \,
\th(-p_1^+) \th(p_2^+) \, Tr M_{2ij}(p_1,p_2) \, \de_{q,-(p_1-p_2)} \nn
\; , \leea{d16}
that can be rewritten after the change of coordinates according to
\bea
p_1 &=& -k \nn \\
p_2 &=& -(k-q) \nn \\
p_2 - p_1 &=& q
\leea{d17}
in the following way
\be
\frac{1}{q^+}\int_k(\et_{k,k-q}g_{k-q,k}-\et_{k-q,k}g_{k,k-q}) \,
\th(k^+)\th(q^+-k^+) \, Tr M_{2ij}(k, k - q)
\; , \lee{d18}
where the symmetry
\bea
\et_{-p_1,-p_2} &=& -\et_{p_1,p_2} \nn\\
g_{-p_1,-p_2} &=& g_{p_1,p_2}
\leea{d19}
has been used. The integration of the commutator over $l$ in the flow equation 
gives rise to 
\bea
&&\hspace{-1em} (\de q_{1\la}^- - \de q_{1\La}^-) \de^{ij}
= \frac{1}{q^+} e^2 \int \frac{d^2k^{\bot}dk^+}{2(2\pi)^3}
\th(k^+) \th(q^+ - k^+) \, \frac{(-1)}{q^- -k^- - (q-k)^-} \\
&&\hspace{2em} \times \, Tr \left( \Ga^i(k,k-q,q) \Ga^j(k-q,k,-q) \right) \,
\left[ \exp\left\{ -2 \left( \frac{\De_{q,q-k}}{\la} \right)^2 \right\}
     - \exp\left\{ -2 \left( \frac{\De_{q,q-k}}{\La} \right)^2 \right\} \right] \nn
. \leea{d20}
This means for the photon energy correction
\bea
\de q_{1\la}^-\de^{ij} &=&
\frac{1}{q^+}e^2 \int \frac{d^2k^{\bot}dk^+}{2(2\pi)^3}
\th(k^+) \th(q^+ - k^+) \\ 
&& \times \, Tr \left( \Ga^i(k,k-q,q) \Ga^j(k-q,k,-q) \right)
\frac{1}{q^- -k^- - (q-k)^-} \, \times(-R) \nn
\; , \leea{d21}
where the regulator $R$
\be
R=\exp\left\{ -2 \left( \frac{\De_{q,k}}{\la} \right)^2 \right\}
\lee{d22}
has been introduced. Define the new set of coordinates
\bea
\frac{(q-k)^+}{q^+} &=& x \nn \\
k &=& ((1-x)q^+,(1-x)q^{\bot}+\kappa^{\bot}) \nn \\
q - k &=& (xq^+,xq^{\bot} - \kappa^{\bot})
\; , \leea{d23}
where $q=(q^+,q^{\bot})$ is the photon momentum.
Then the terms present in $\de q_{1\la}^-$ are
\bea
& \Ga^i(k,k-q,q)\Ga^i(k-q,k,-q)
=\frac{2}{(q^+)^2x(1-x)^2}
\left( \left( 2x-2+\frac{1}{x} \right) \kappa^{\bot 2} + \frac{m^2}{x} \right)
 & \nn \\
& \De_{k-q,k} = q^- - k^- - (q - k)^- =
-\frac{\kappa^{\bot 2} + m^2}{q^+x(1-x)}+ \frac{q^2}{q^+} = 
\frac{\tilde{\De}_{k-q,k}}{q^+} &
\; . \leea{d24}
The integral for the photon energy correction corresponding
to the first diagram of \fig{photselfen} takes the form 
\bea
q^+\de q_{1\la}^- &\hspace{-3mm}=\hspace{-3mm}&
 -\frac{e^2}{8\pi^2}\int_0^1 \!\! dx 
 \int d\kappa_{\bot}^2
 \frac{(2x-2+\frac{1}{x})\kappa_{\bot}^2+\frac{m^2}{x}}
 {(1-x)(\kappa_{\bot}^2+f(x))} \!\!\times\! (-R) \\
&\hspace{-3mm}=\hspace{-3mm}&-\frac{e^2}{8\pi^2}\int_0^1 \!\! dx
 \int d\kappa_{\bot}^2
 \left\{ \frac{q^2}{\kappa_{\bot}^2+f(x)}
  \left( 2x^2-2x+1 \right) +\frac{2m^2}{[1-x]}
  + \left( -2+\frac{1}{[x][1-x]} \right) \right\} \!\!\times\! (-R) \nn
\leea{d25}
with
\be
f(x)=m^2-q^2x(1-x)
\; , \lee{d26}
and the principal value prescription, denoted by '\mbox{\boldmath{$[\;]$}}',
introduced to regularize the IR divergencies.

This is the form of the photon correction used in the main text.

\newpage
%FiguresI
\newpage

\newpage
% FiguresIV

%\vspace*{1cm}

% FiguresII

%\vspace*{1cm}

% FiguresIII

\end{document}